\documentclass[twocolumn,showkeys,showpacs]{revtex4-2}
\usepackage[cp1251]{inputenc}
\usepackage{amsmath}
\usepackage{amsfonts}
\usepackage{amssymb}
\usepackage{graphicx}
\usepackage{textcomp}
\usepackage{color}
\begin{document}

\title{Magnetization reversal of finite-length Co and Fe atomic chains on Pt(332) surface: numerical calculations and a new theoretical approach}

\author{S.V. Kolesnikov}
\email{kolesnikov@physics.msu.ru}
\author{E.S. Glazova}
\author{A.M. Saletsky}
\affiliation{Faculty of Physics, Lomonosov Moscow State University, Moscow 119991, Russian Federation}

\begin{abstract}
Different mechanisms of magnetization reversal in finite-length Co and Fe chains on the Pt(332) surface have been investigated, taking into account the Dzyaloshinskii-Moriya interaction.
It has been found that the magnetization reversal in short atomic chains occurs through the simultaneous reversal of all magnetic moments.
In contrast, the magnetization reversal in long atomic chains is facilitated by the formation of domain walls, which exhibit distinct structures for Co and Fe atomic chains.
Using the geodesic nudged elastic band method, we have determined the energy barriers for magnetization reversal in chains consisting of 5 to 100 atoms.
Additionally, the frequency prefactors have been calculated within the framework of the harmonic approximation of transition state theory.
Notably, the dependencies of these prefactors on chain length and external magnetic field are significant and non-monotonic.
We propose a theoretical approach that qualitatively describes the numerical dependencies for both the energy barriers and the frequency prefactors.
The magnetization curves derived from our theoretical estimates show qualitative agreement with the results of numerical calculations.
This analytical approach enables the estimation of the coercive force of atomic chains across a wide range of lengths, temperatures, sweeping rates, and model parameters.
The proposed theoretical framework is applicable not only to the Co and Fe chains on the Pt(332) surface but also to a broad class of one-dimensional magnetic systems.
\end{abstract}	


\keywords{atomic chains, magnetic properties, geodesic nudged elastic band method, transition state theory}

\date{\today}
	
\maketitle

\section{Introduction}

A theoretical investigation into the magnetic properties of atomic chains is particularly appealing due to their potential applications in
mass storage devices~\cite{JPCM16.R603,JPCM22.433001,NANO6.1},
spintronics~\cite{RMP76.323},
quantum computing~\cite{Mermin_book},
quantum sensing~\cite{shukla2024},
quantum communications~\cite{PRL91.207901,EPL119.30001,QIP20.54,AP442.168918},
and other fields~\cite{RMP91.041001,UFN64.671}.
An intensive study of magnetic properties of single-atomic chains has begun after the discovery of the giant magnetic anisotropy energy (MAE) of Co atoms and Co atomic chains on the Pt(997) surface~\cite{Nature416.301,Science300.1130}
using X-ray magnetic circular dichroism and scanning tunneling microscope~\cite{BRUNE20091812,BANSMANN2005189,PhysRevLett.102.257203}.
It is important that ferromagnetic Co chains can grow on the step edges of the Pt(997) surface as a result of a self-organization process occurring at low concentrations of Co atoms and low temperatures.
A similar phenomenon has been experimentally observed for ferromagnetic Fe atomic chains on the Cu(111) surface~\cite{JPCM15.R1,PRB56.2340}.
To enhance information recording density, biatomic ferromagnetic chains~\cite{PRL93.077203,NJPhys17.023014} and antiferromagnetic chains~\cite{NuturePhys14.213,RevModPhys90.015005,NatureNanotech11.231} can be employed as bits of information.
In practice, the maximum length of homogeneous atomic chains does not exceed 100 atoms~\cite{RMP91.041001,Nature416.301,UFN64.671,ASS404.12}.
The structure, magnetic properties, and thermodynamic characteristics of such atomic chains are significantly influenced by their length~\cite{UFN64.671,Science300.1130,SYROMYATNIKOV201669,PhysRevB.72.224410}.

The magnetic properties of atomic chains are typically described within the framework of an effective theory that includes exchange interaction, magnetic anisotropy energy (MAE), dipole-dipole interaction (DDI), and interaction with external magnetic fields~\cite{RMP91.041001,UFN64.671,PRB68.104436,NatCommun10.2565}.
In the case of Co and Fe atomic chains on a platinum surface, the Dzyaloshinskii-Moriya interaction~\cite{JPCS4.241,PRL4.228} (DMI) should also be considered.
Historically, DMI was proposed by Igor Dzyaloshinskii to explain the weak ferromagnetism observed in antiferromagnetic crystals $\alpha{\rm-Fe}_2{\rm O}_3$, ${\rm MnCO}_3$, and ${\rm CoCO}_3$~\cite{JPCS4.241}.
However, interest in DMI has drastically arisen following the observation of strong DMI in transition metal films~\cite{NatPhys11.825,NatCommun6.7635}.
In layered structures, DMI plays a significant role in the magnetic properties of chiral domain walls, chiral bubbles, and skyrmions~\cite{Fert2017,Soumyanarayanan2016,PhysRevLett.129.126101,PhysRevB.107.L100419,PhysRevB.109.214510}.
Strong DMI in atomic chains also can lead to some interesting phenomena.
For instance, an atomic chain may exhibit a DMI-induced non-collinear magnetic ground state, as seen in Fe chains on the Ir(001) surface~\cite{PRB80.195420,PRB.79.134402,PRL.108.197204}.
In finite-length atomic chains, DMI can induce non-collinearity at the ends of the chain, even when the magnetic moments are collinear in the middle of the chain~\cite{PRB88.184422,KOLESNIKOV2023170869}.
As a result, DMI can significantly modify the energies of both the ground and excited states of atomic chains~\cite{PRB.78.140403,PRB94.024403,PRB106.014417,JNN11.3005}.
In certain scenarios, the interplay between MAE and DMI can result in the formation of exotic domain walls, which represent intermediate configurations between Bloch and N\'eel walls~\cite{PRL127.127203,JMMM409.155,Kolesnikov_EPJB2023}.

The parameters of an effective Hamiltonian for atomic chains can be calculated from first principles~\cite{RMP71.1253,RPP74.096501}.
Unfortunately, {\it ab initio} methods are applicable only in two limiting cases: either for very short atomic chains~\cite{PhysRevLett.101.107204,JPCM19.446001} or for infinitely long ones~\cite{PRB94.024403,Tsysar_2015}.
The calculated parameters can then be utilized for further investigations into the magnetic properties of finite-length atomic chains through solutions of the Landau-Lifshitz-Gilbert equation~\cite{Landau,Evans_2014} or kinetic Monte Carlo (kMC) simulations~\cite{PRB73.174418,EPL137.56003}.
At low temperatures, the time-consuming kMC simulations can be substituted with analytical estimations~\cite{JETPL103.588,JETP125.644,PRB100.224424}.
If atomic chains are not very long, then the magnetization reversal frequencies can be computed using transition state theory (TST)~\cite{RMP62.251,Coffey.ch5}.
In this context, it is crucial to accurately calculate the energy barriers between the ground and excited states of the chain.
A powerful method for numerically calculating these energy barriers in magnetic systems is the geodesic nudged elastic band (GNEB) method~\cite{CPC196.335,JETPL113.801}.
From experimental data, the frequency prefactor can be estimated to an order of magnitude~\cite{Nature416.301}.
For more precise calculations, the harmonic approximation of TST~\cite{JETPL113.801,ZPC227.1543,PhysRevB.85.184409,PhysRevLett.110.020604} can be employed.
The energies of the ground and excited states can be estimated using a continuous model~\cite{PRB.78.140403,PhysRevLett.65.787,PhysRevB.50.16485,PhysRevB.101.184424}.
In particular, the energy barriers for magnetization reversal in finite-length Co and Fe chains have been recently calculated in the absence of an external magnetic field~\cite{KOLESNIKOV2023170869,Kolesnikov_EPJB2023}.
However, the analytical calculation of frequency prefactors presents a more challenging problem, which can be solved only in certain simple cases for infinitely long atomic chains~\cite{PhysRev.130.1677,PhysRevB.50.16501}.

The main goal of this paper is the numerical and analytical investigations into the ground and excited states of finite-length Co and Fe chains on the Pt(332) surface.
Specifically, we aim to address the following issues:
(i) we will generalize the methodology for calculating energy barriers in the presence of a non-zero external magnetic field;
(ii) we will calculate and discuss, for the first time, the frequency prefactors associated with the magnetization reversal of finite-length Co and Fe chains;
(iii) we will develop an analytical method for estimating these prefactors, which will yield qualitative agreement with the numerical results;
and (iv) utilizing our method, we will estimate the coercive force of the atomic chains across a broad range of lengths, temperatures, sweeping rates, and model parameters.
The results presented in this paper will be of interest to specialists in both theoretical and experimental physics related to one-dimensional magnetism.

The rest of the paper is organized as follows.
The methodology Section~\ref{sec:methods} is divided into two parts: the theoretical model is discussed in Section~\ref{sec:methods_Hamiltonian}, while the computational methods, including transition state theory (TST) and the geodesic nudged elastic band (GNEB) method, are briefly outlined in Section~\ref{sec:methods_TST}.
The main results of our investigation, presented in Section~\ref{sec:RandD}, are divided into four parts.
The numerical results from GNEB and TST are detailed in Section~\ref{sec:numerical_results}.
Sections~\ref{sec:analytical_barriers} and~\ref{sec:analytical_prefactors} develop the analytical methods for estimating the energy barriers and frequency prefactors, respectively.
The magnetization properties of finite-length Co and Fe chains, across a wide range of lengths, temperatures, sweeping rates, and model parameters, are discussed in Section~\ref{sec:magnetization_properties}.
Finally, Section~\ref{sec:conclusion} concludes the paper.

\section{Methods}\label{sec:methods}

\subsection{The Hamiltonian}\label{sec:methods_Hamiltonian}

The magnetic properties of finite-length atomic chains can be adequately described within the framework of the generalized Heisenberg model.
At very low temperatures, the quantum tunneling effect should be considered~\cite{RMP62.251,PRL60.661}.
However, this effect can be neglected at temperatures exceeding 1 K~\cite{PRL95.237203,NJP11.063004}, allowing for the use of classical unit vectors ${\bf s}_i$ (${\bf s}_i^2=1$) in place of quantum operators.

\begin{figure}[htb]
\begin{center}
\includegraphics[width=0.6\linewidth]{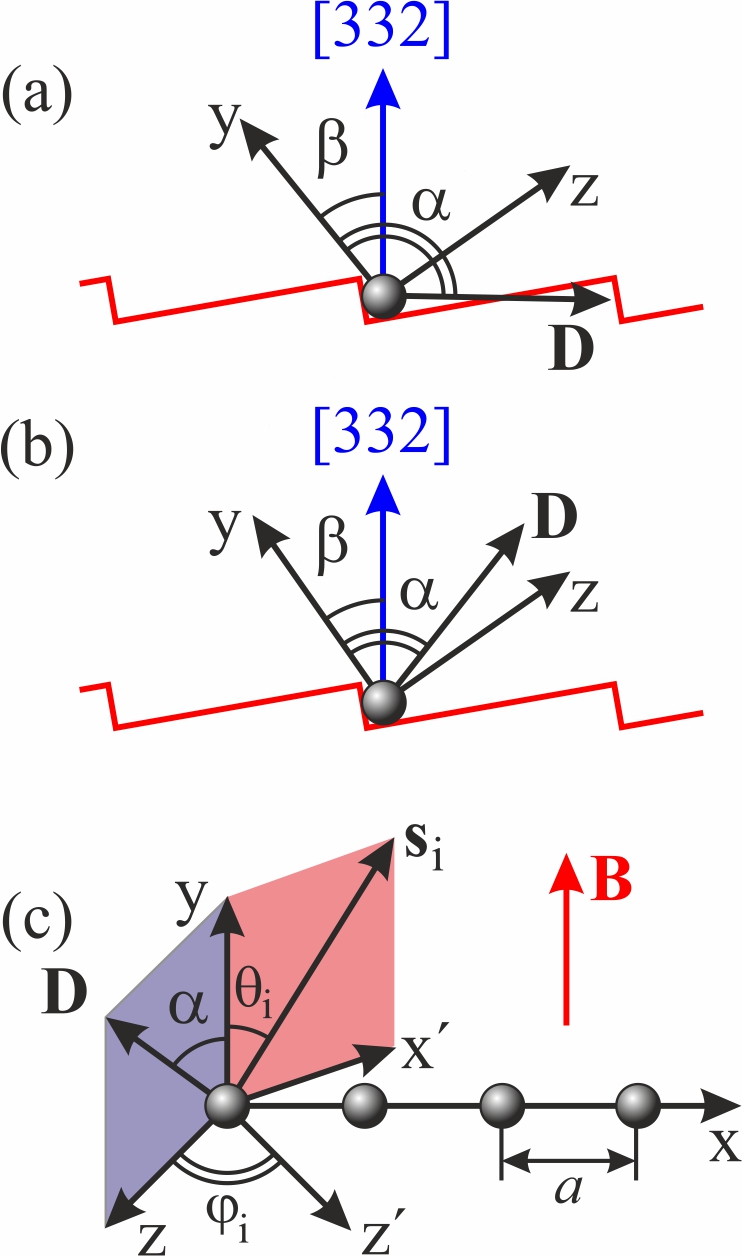}
\caption{\label{fig1} The relative orientations of the magnetization axes, the Dzyaloshinskii vector, and the Pt(332) surface are shown for the cases of Co chains (a) and Fe chains (b).
(c) A schematic view of the atomic chain.
The vector ${\bf D}$ is positioned in the $YZ$-plane, and the unit vector ${\bf s}_i$ lies in the $X^\prime Y$-plane.}
\end{center}
\end{figure}

The magnetic interaction of Co or Fe atoms on the Pt(332) surface can be described by the effective Hamiltonian~\cite{PRB94.024403,RMP91.041001}
\begin{equation}\label{eq:eff_ham}
H=H_{\rm ex}+H_{\rm MAE}+H_{\rm int}+H_{\rm dip},
\end{equation}
where the first term describes the exchange interaction between the atoms in the chain.
Following Ref.~\cite{PRB94.024403}, we neglect the symmetric anisotropic exchange interaction.
Thus, the term $H_{\rm ex}$ can be separated into two contributions~\cite{RMP91.041001}
\begin{equation}\label{eq:exchange}
H_{\rm ex}=-J\sum_{i}\left({\bf s}_i\cdot{\bf s}_{i+1}\right)-{\bf D}\sum_{i}\left[{\bf s}_i\times{\bf s}_{i+1}\right].
\end{equation}
The first term in Eq. (\ref{eq:exchange}) represents the standard Heisenberg Hamiltonian.
In our model, we assume that
(i) only the exchange interaction between nearest neighbors is considered,
(ii) the exchange integrals $J$ between all nearest neighbors in the chain are the same,
and (iii) the chain exhibits ferromagnetic ordering below the critical temperature ($J>0$).
The second term in Eq. (\ref{eq:exchange}) represents DMI~\cite{JPCS4.241,PRL4.228}.
In infinite chains, the Dzyaloshinskii vector ${\bf D}$ is oriented in a plane perpendicular to the atomic chain~\cite{PRB94.024403} and is consistent for all atoms (see Fig.~\ref{fig1}).
We further assume that these characteristics of the vector ${\bf D}$ hold true for finite-length chains as well.

The second term in Hamiltonian (\ref{eq:eff_ham}) represents MAE and can be expressed as follows~\cite{RMP91.041001}
\begin{equation}\label{eq:MAE}
H_{\rm MAE}=\sum_{i}\left[-K(s_i^y)^2+E\left((s_i^z)^2-(s_i^x)^2\right)\right].
\end{equation}
We assume that the parameter $K$ is positive and that $K>|E|$, indicating that the $y$-axis is the easy axis of magnetization.
The parameter $E$ is positive for Co chains and negative for Fe chains on the Pt(332) surface (see Table~\ref{table}).
Thus, the $z$-axis is the hard axis of magnetization for the Co chain, while the $x$-axis is the hard axis for the Fe chain.
As will be discussed below, this distinction between the two systems results in different domain wall structures in long Co and Fe chains.
The relative orientations of the $y$-axis, $z$-axis, the Dzyaloshinskii vector ${\bf D}$, and the Pt(332) surface are illustrated in Figures~\ref{fig1}(a) and~\ref{fig1}(b) for the Co and Fe chains, respectively.
The values of angles $\alpha$ and $\beta$ are provided in Table 1~\ref{table}.
The $x$-axis is aligned along the atomic chain (see Fig.~\ref{fig1}(c)).

The third term in Hamiltonian (\ref{eq:eff_ham}) represents the Zeeman interaction with the external magnetic field ${\bf B}$
\begin{equation}\label{eq:Zeeman}
H_{\rm int}=-\mu\sum_{i}\left({\bf s}_i\cdot{\bf B}\right),
\end{equation}
where $\mu$ denotes the magnetic moment of the Co or Fe atoms.
We assume that the values of $\mu$ are the same for all atoms in the chain and that $\mu$ is independent of the direction of the unit vector ${\bf s}_i$.

The last term in the Hamiltonian (\ref{eq:eff_ham}) represents DDI
\begin{equation}\label{eq:dip-dip}
H_{\rm dip}=\frac{\mu_0\mu^2}{4\pi}\sum\limits_{i>j}\frac{\left({\bf s}_i\cdot{\bf s}_j\right)r_{ij}^2- 3\left({\bf s}_i\cdot{\bf r}_{ij}\right)\left({\bf s}_j\cdot{\bf r}_{ij}\right)}{r_{ij}^5},
\end{equation}
where ${\bf r}_{ij}={\bf r}_{i}-{\bf r}_{j}$ is the radius vector between the $i$-th and $j$-th atoms, $\mu_0/4\pi=10^{-7}$~H/m.

The Hamiltonian parameters calculated for infinite Co and Fe chains on the Pt(332) surface using DFT~\cite{PRB94.024403} are presented in Table~\ref{table}.
The magnetic moment $\mu$ is the sum of the spin magnetic moment $\mu_{S}$ and the orbital magnetic moment $\mu_{L}$.
The spin magnetic moment $\mu_{S}$ is $2.20\mu_{\rm B}$ for Co atoms and $3.27\mu_{\rm B}$ for Fe atoms.
The orbital magnetic moment $\mu_{L}$ is $0.19\mu_{\rm B}$ for Co atoms and $0.14\mu_{\rm B}$ for Fe atoms.
The corresponding spin and orbital quantum numbers are $S=1.1$ and $L=0.19$ for Co atoms, and $S=1.635$ and $L=0.14$ for Fe atoms.
Assuming the total angular momentum quantum number is $J=L+S$, we can calculate the Land\'e $g$-factor, which is also presented in Table~\ref{table}.
The nearest neighbor distance $a$ is 2.82~\r{A}.
This set of parameters will be used for calculations in the case of finite-length atomic chains~\footnote{The angle $\beta$ is not required for the calculations.}.
It means that we neglect possible deviations in the magnetic properties of the edge atoms~\cite{PhysRevB.68.024433,JPCM19.446001}.
Considering these deviations results in a very weak effect for long atomic chains~\cite{JETP125.644}.

\begin{table}[!h]
\caption{The parameters of the Hamiltonian (\ref{eq:eff_ham}) according to DFT calculations~\cite{PRB94.024403}.
The parameters $J$, $K$, $E$ and $D$ are in meV, $\mu$ is in Bohr magneton ($\mu_{\rm B}$) units, the angles $\alpha$ and $\beta$ are in degrees.}
\label{table}
\begin{center}
\begin{tabular}{@{}ccccccccc}
\hline
\hline
Atom & $J$ & $K$ & $\beta$ & $E$ & $D$ & $\alpha$ & $\mu$ & $g$ \\
\hline
Co & ~61.8~ & ~1.31~ & ~$29^\circ$~ & ~0.34~ & ~1.92~ & ~$131^\circ$~ & ~2.39~ & ~1.85~ \\
Fe & ~45.9~ & ~1.73~ & ~$25^\circ$~ & -0.89~ & ~1.98~ & ~$74^\circ$~ & ~3.41~ & ~1.92~ \\
\hline
\hline
\end{tabular}
\end{center}
\end{table}

Previous investigations~\cite{KOLESNIKOV2023170869,9969869,Kolesnikov_EPJB2023} have demonstrated that Co and Fe chains exhibit quasi-collinear ground states in the absence of an external magnetic field.
It means that the magnetic moments in the middle of the atomic chain are almost collinear, while the magnetic moments at the ends of the chain rotate anticlockwise in the $XY$-plane as the atomic number $i$ increases.
Transitions between these two quasi-collinear states can occur through two distinct mechanisms.
The first mechanism involves the reversal of magnetic moments without the formation of domain walls, which is more energetically favorable for short atomic chains.
The second mechanism involves the transition through the formation of a domain wall, which is preferred for long atomic chains.
In other words, short chains behave like superparamagnetic particles, while longer chains exhibit ferromagnetic behaviour.
It is important to note that this description holds true only when the temperature is below the critical temperature $T_{\rm cr}$ of the atomic chain.
To estimate the value of $T_{\rm cr}$, we performed a series of equilibrium Monte Carlo simulations~\cite{Frenkel_Smit_book,Kamilov:1999}.
According to these simulations, the critical temperature for the Co/Pt(332) chain  consisting of 100 atoms is found to be $40\pm2$~K, while the corresponding critical temperature for the analogous Fe chain is $35\pm2$~K.
It should be noted that more precise calculations of the critical temperature are beyond the scope of this study.
For our purposes, it is important that the presented model remains valid in the temperature range of 1 K to 30 K.

\subsection{The transition state theory}\label{sec:methods_TST}

In order to estimate the magnetization reversal frequencies $\nu$ of the atomic chains we employ the transition state theory (TST)~\cite{RMP62.251,JETPL113.801}
\begin{equation}\label{eq:frequency}
\nu=\nu_0\exp\left(-\frac{\Delta E}{k_B T}\right),
\end{equation}
where $T$ is the temperature of the system, $k_B$ is the Boltzmann constant, $\Delta E$ is the energy barrier between two states, and $\nu_0$ is the frequency prefactor.
It is assumed that the values of $\Delta E$ do not depend on temperature~\cite{RMP62.251}.
GNEB method~\cite{JETPL113.801,CPC196.335} is utilized to calculate the energy barriers $\Delta E$.
Generally, the GNEB method does not guarantee the finding of all possible energy barriers.
The success of GNEB calculations relies on the selection of the initial state, the final state, and the initial path connecting them.
However, the finite-length Co and Fe atomic chains are relatively simple systems.
Thus, the energy barriers $\Delta E$ can be accurately calculated using the GNEB method.

In the harmonic approximation of TST, the frequency prefactor $\nu_0$ is calculated as follows~\cite{JETPL113.801,ZPC227.1543,PhysRevB.85.184409,PhysRevLett.110.020604}:
\begin{equation}\label{eq:prefactor}
\nu_0=\frac{1}{2\pi}\left[\left(\sum_{k=1}^{2N-1}\frac{(a_{\rm sp}^{k})^2}{\xi_{\rm sp}^{k}}\right)
\frac{\prod\limits_{k=1}^{2N}\xi_{\rm min}^{k}}{\prod\limits_{k=1}^{2N-1}\xi_{\rm sp}^{k}} \right]^{1/2},
\end{equation}
where $\xi_{\rm min}^{k}$ and $\xi_{\rm sp}^{k}$ are the eigenvalues of the $2N$-dimensional Hessian matrix $\mathcal{H}$ at the local energy minimum and the saddle point, respectively.
We assume that the values of $\xi_{\rm min}^{k}$ and $\xi_{\rm sp}^{k}$ are arranged in descending order:
$\xi_{\rm min}^{1}\ge\xi_{\rm min}^{2}\ge\ldots\ge\xi_{\rm min}^{2N}>0$ and $\xi_{\rm sp}^{1}\ge\xi_{\rm sp}^{2}\ge\ldots\ge\xi_{\rm sp}^{2N-1}>0$, while $\xi_{\rm sp}^{2N}<0$.
The coefficients $a_{\rm sp}^{k}$ are calculated as follows~\cite{JETPL113.801}
\begin{equation}\label{eq:prefactor_a}
a_{\rm sp}^{k}=\frac{\gamma}{\mu}\left(e_{\rm sp}^{2N}\left[s_{\rm sp}\times e_{\rm sp}^{k}\right]\right),
\end{equation}
where $e_{\rm sp}^{k}$ are eigenvectors of the Hessian matrix of energy at the saddle point $s_{\rm sp}=({\bf s}_{{\rm sp},1},\ldots,{\bf s}_{{\rm sp},N})$,
$\gamma=g\mu_B/\hbar$ is the gyromagnetic ratio.
It is important to emphasize that we consider finite-length atomic chains in the presence of biaxial magnetocrystalline anisotropy (\ref{eq:MAE}).
The Hamiltonian (\ref{eq:eff_ham}) does not exhibit invariance under translations or rotations.
As a result, all eigenvalues $\xi_{\rm min}^{k}$ and $\xi_{\rm sp}^{k}$ of the Hessian matrix $\mathcal{H}$ are non-zero.

\section{Results and Discussions}\label{sec:RandD}

Due to technical limitations, the creation of homogeneous atomic chains consisting of more than 100 atoms of Co or Fe is nearly impossible~\cite{RMP91.041001,Nature416.301,UFN64.671,ASS404.12}.
Therefore, in this Section, we will focus on the magnetic properties of atomic chains consisting of $N\le100$ atoms.
First of all, we will present the numerical results obtained from GNEB method and TST.
To provide a physical interpretation of these numerical results, we will discuss them within the framework of simple theoretical models.
Finally, we will compare the magnetization properties of finite-length chains of Co and Fe, based on both the numerical results and theoretical estimates.

\subsection{Numerical results}\label{sec:numerical_results}

\begin{figure}[htb]
\begin{center}
\includegraphics[width=1.0\linewidth]{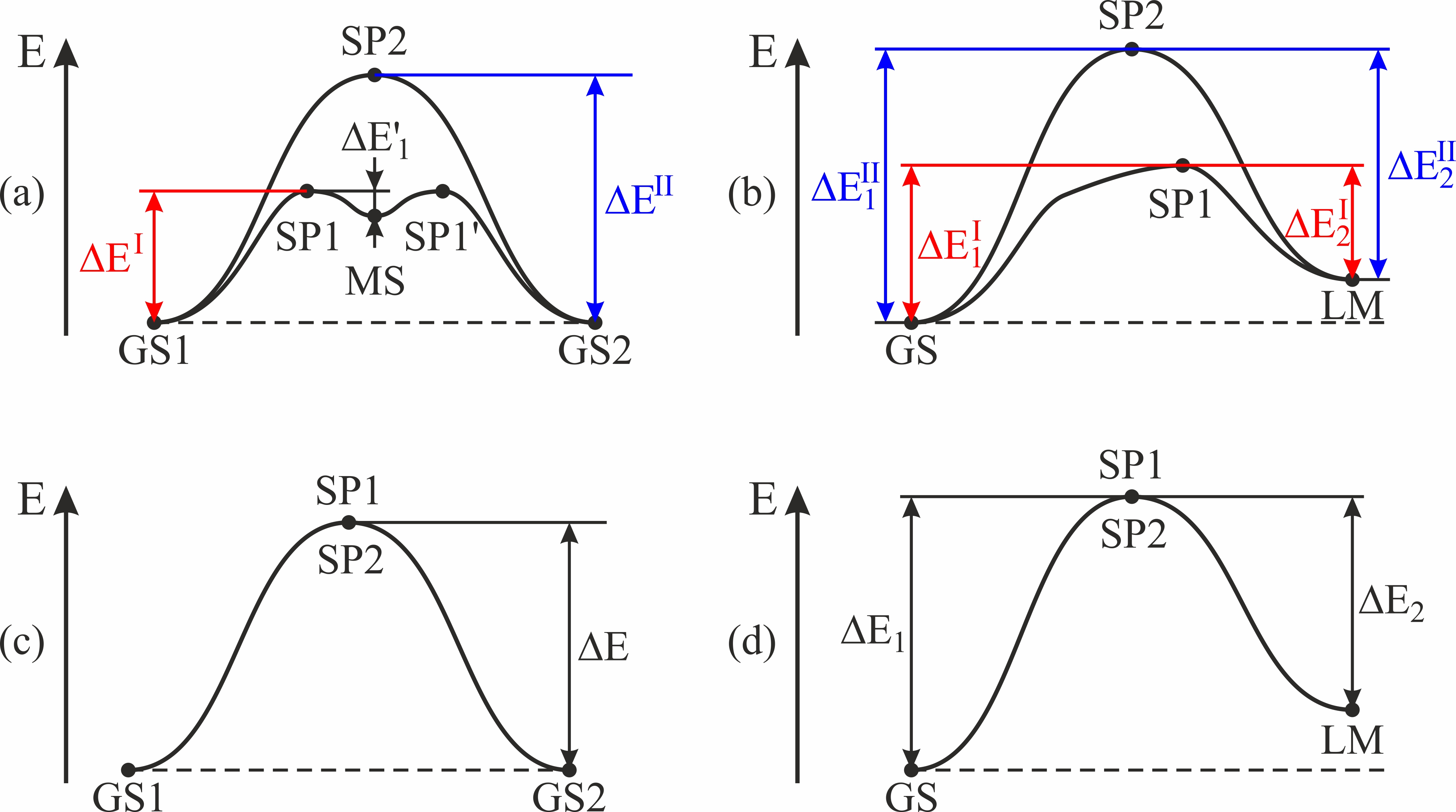}
\caption{\label{fig2} The schematic energy diagrams illustrate the states of (a) the Co chain at $B_y=0$, (b) the Co chain at $B_y\neq0$, (c) the Fe chain at $B_y=0$, and (d) the Fe chain at $B_y\neq0$.}
\end{center}
\end{figure}

First, let us briefly summarize the main results previously obtained for Co~\cite{KOLESNIKOV2023170869} and Fe~\cite{Kolesnikov_EPJB2023} finite-length atomic chains in the absence of an external magnetic field (${\bf B}=0$).
It has been found that DMI leads to several common effects in both Co and Fe chains.
(i) In the ground state, the magnetic moments in the atomic chain are quasi-collinear.
While the magnetic moments in the middle of the chain are nearly collinear, those at the ends rotate anticlockwise in the $XY$-plane as the atomic index $i$ increases.
(ii) Magnetization reversal in short atomic chains occurs without the formation of domain walls.
(iii) In contrast, magnetization reversal in long atomic chains happens via the formation of domain walls.
(iv) DDI has a weak influence on the energy barrier values.
The differences between Co and Fe chains are the following.
In Co atomic chains, the domain walls are N\'eel walls.
These domain walls can be either clockwise (CDW, corresponding to SP2 state in the energy diagram in Fig.~\ref{fig2}(a)) or anticlockwise  (ACDW, corresponding to SP1 and SP$1^\prime$ states).
The formation of ACDW is energetically favorable.
If magnetization reversal occurs through the formation of ACDW, an additional metastable state (MS) may emerge.
This metastable state corresponds to the positioning of the ACDW in the middle of the atomic Co chain.
At the same time, in Fe chains, the interplay between MAE and DMI results in a rotation of the domain wall plane, yielding domain walls that are intermediate configurations between Bloch and N\'eel walls.
In the case of Fe chains, we can also identify two types of domain walls.
However, their energy levels are nearly identical (as seen in the energy diagram in Fig.~\ref{fig2}(c)).
Notably,  there are no metastable states during the magnetization reversal in the Fe chains.

Let us discuss how this scenario changes in the presence of a non-zero external magnetic field ${\bf B}$.
We assume that a uniform magnetic field is applied along the easy axis of magnetization (${\bf B}=B_y{\bf e}_y$).
In this case, the states GS1 (where ${\bf s}_i$ are aligned with ${\bf B}$) and GS2 (where ${\bf s}_i$ are aligned against ${\bf B}$) become nonequivalent.
The state GS1 transforms into a non-degenerate ground state (GS), while the state GS2  transforms to a local minimum (LM) of the energy (refer to the energy diagrams in Fig.~\ref{fig2}(b,d)).
For the Co atomic chains, the metastable state (MS) disappears when $|B_y|>0.03$~T.
We will consider the magnetic field $B_y$ in the range from -5 T to 5 T, allowing us to neglect this metastable state.

\begin{figure*}[htb]
\begin{center}
\includegraphics[width=0.9\linewidth]{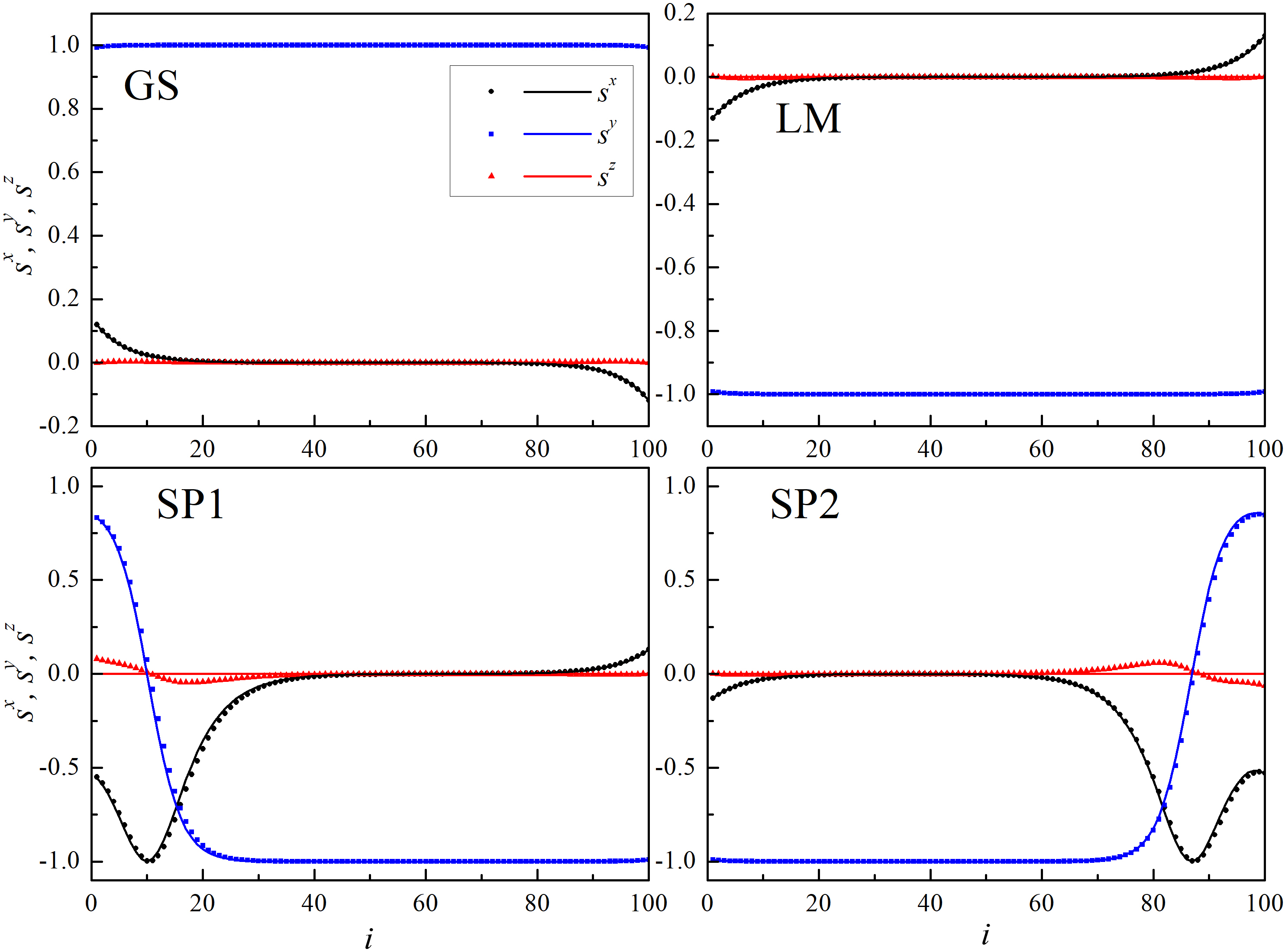}
\caption{\label{fig3} The magnetic configurations of a Co/Pt(332) chain consisting of $N=100$ atoms corresponding to different states: the ground state (GS), the local minimum (LM), the low-energy saddle point (SP1), and the high-energy saddle point (SP2).
The external magnetic field is set to $B_y=1$~T.
The points represent the results of the GNEB calculations.
The solid lines show the theoretical approximations.}
\end{center}
\end{figure*}

Figure~\ref{fig3} illustrates the magnetic configurations of a Co/Pt(332) chain consisting of $N=100$ atoms at $B_y=1$~T.
The points in the figure represent the results of the numerical GNEB calculations.
In both the GS and LM states, the magnetic moments are quasi-collinear, similar to the situation observed at $B_y=0$.
The presence of DMI causes an anticlockwise rotation of the magnetic moments at the ends of the chain~\cite{PRB88.184422,KOLESNIKOV2023170869}.
However, the maximal angles $\theta_\text{end}$ between the magnetic moments and the $y$-axis differ for the GS and LM states, with the condition $\theta_\text{end}^\text{LM}>\theta_\text{end}^\text{GS}$.
The transition between GS and LM states occurs through the formation of a N\'eel domain wall.
The deviations of the magnetic moments from the $XY$-plane are minimal.
As observed in the case of $B_y=0$, the DMI leads to an energy disparity between CDW and ACDW, with $E_\text{CDW}>E_\text{ACDW}$.
In Figure~\ref{fig3}, the low-energy saddle point SP1 corresponds to the ACDW, which is positioned close to the left end of the chain.
The high-energy saddle point SP2 corresponds to the CDW, located near the right end of the chain.
The configuration at the opposite end of the chain remains the same as observed in the LM state.

\begin{figure*}[htb]
\begin{center}
\includegraphics[width=0.9\linewidth]{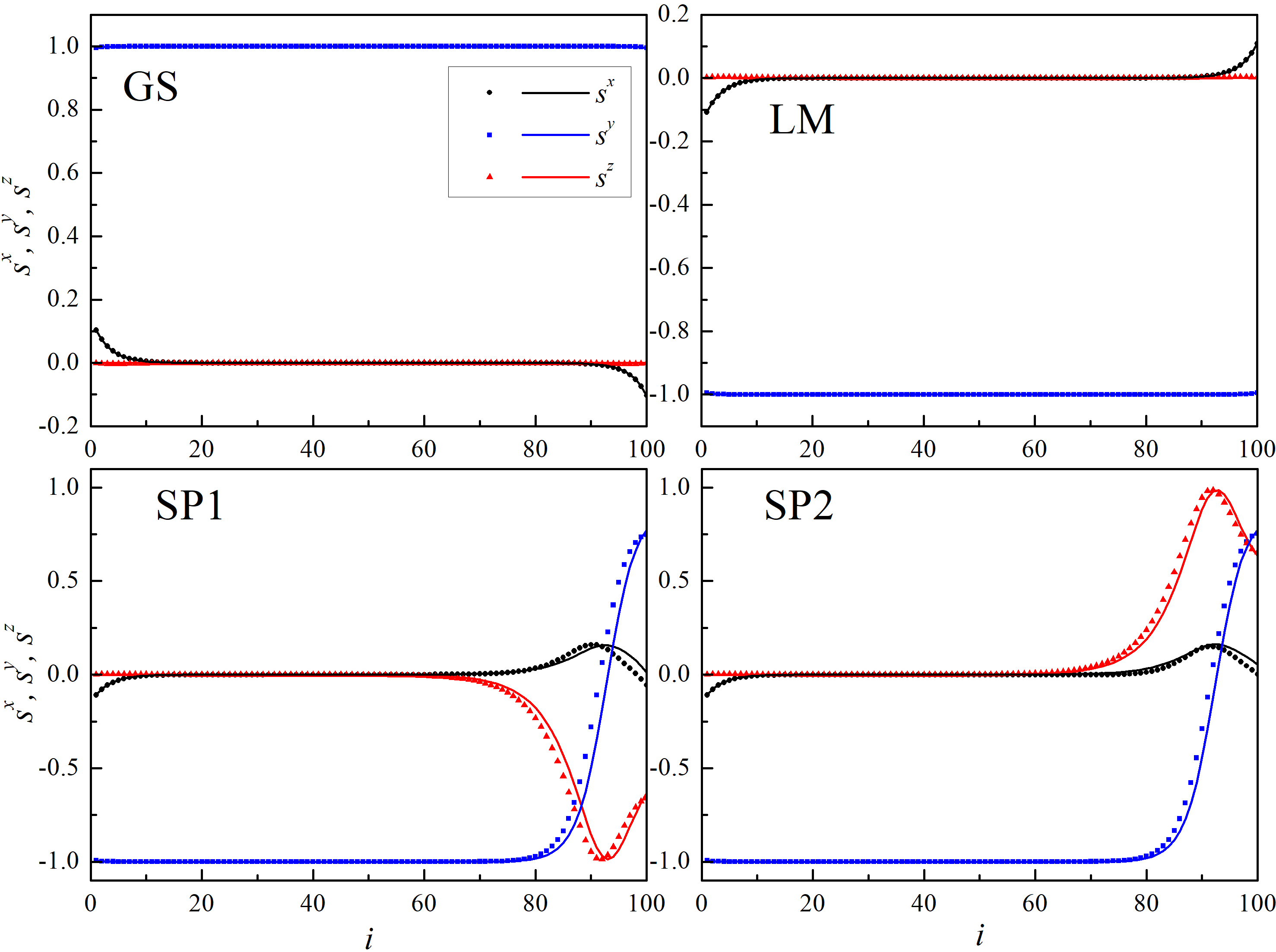}
\caption{\label{fig4} The magnetic configurations of a Fe/Pt(332) chain consisting of $N=100$ atoms corresponding to different states: the ground state (GS), the local minimum (LM), the saddle points (SP1 and SP2).
The external magnetic field is set to $B_y=1$~T.
The points represent the results of the GNEB calculations.
The solid lines show the theoretical approximations.}
\end{center}
\end{figure*}

Figure~\ref{fig4} presents the magnetic configurations of a Fe/Pt(332) chain consisting of $N=100$ atoms under an external magnetic field of $B_y=1$~T.
The points in the figure represent the results of the numerical GNEB calculations.
The magnetic configurations in both the GS and LM states are very similar to those observed in the cobalt chain.
The magnetic moments are quasi-collinear, and DMI induces an anticlockwise rotation of the moments at the ends of the chain~\cite{Kolesnikov_EPJB2023}.
All magnetic moments remain confined to the $XY$-plane.
As with the Co/Pt(332) system, transitions between GS and LM states occur through the formation of domain walls.
However, the structure of the domain walls in the Co/Pt(332) and Fe/Pt(332) systems is fundamentally different.
In the Fe/Pt(332) system, the easy magnetization axis is perpendicular to the chain, and Bloch domain walls are energetically favorable in the absence of DMI.
The presence of DMI, however, rotates the domain wall plane by an angle of $\pi/2-\varphi_0$, where $\varphi_0$ is the angle between the $XY$-plane and the $X^\prime Y$-plane (as illustrated in Figure~\ref{fig1}(c), where $\varphi_i=\varphi_0$ for all magnetic moments in the domain wall).
Figure~\ref{fig4} shows that the finite-length iron chain has two non-equivalent saddle points (SP1 and SP2).
In both states, the domain wall is positioned closer to the right end of the chain, with $|\varphi_0|\approx81^\circ$, and their energies are nearly identical, with $|E_\text{SP1}-E_\text{SP2}|\approx10^{-4}$~meV.
However, the sign of the angle $\varphi_0$ differs: $\varphi_0>0$ in the SP1 state and $\varphi_0<0$ in the SP2 state.
The magnetic moments at the left end of the chain remain in the $XY$-plane and rotate anticlockwise, similar to the configuration observed in the LM state.

\begin{figure*}[htb]
\begin{center}
\includegraphics[width=0.9\linewidth]{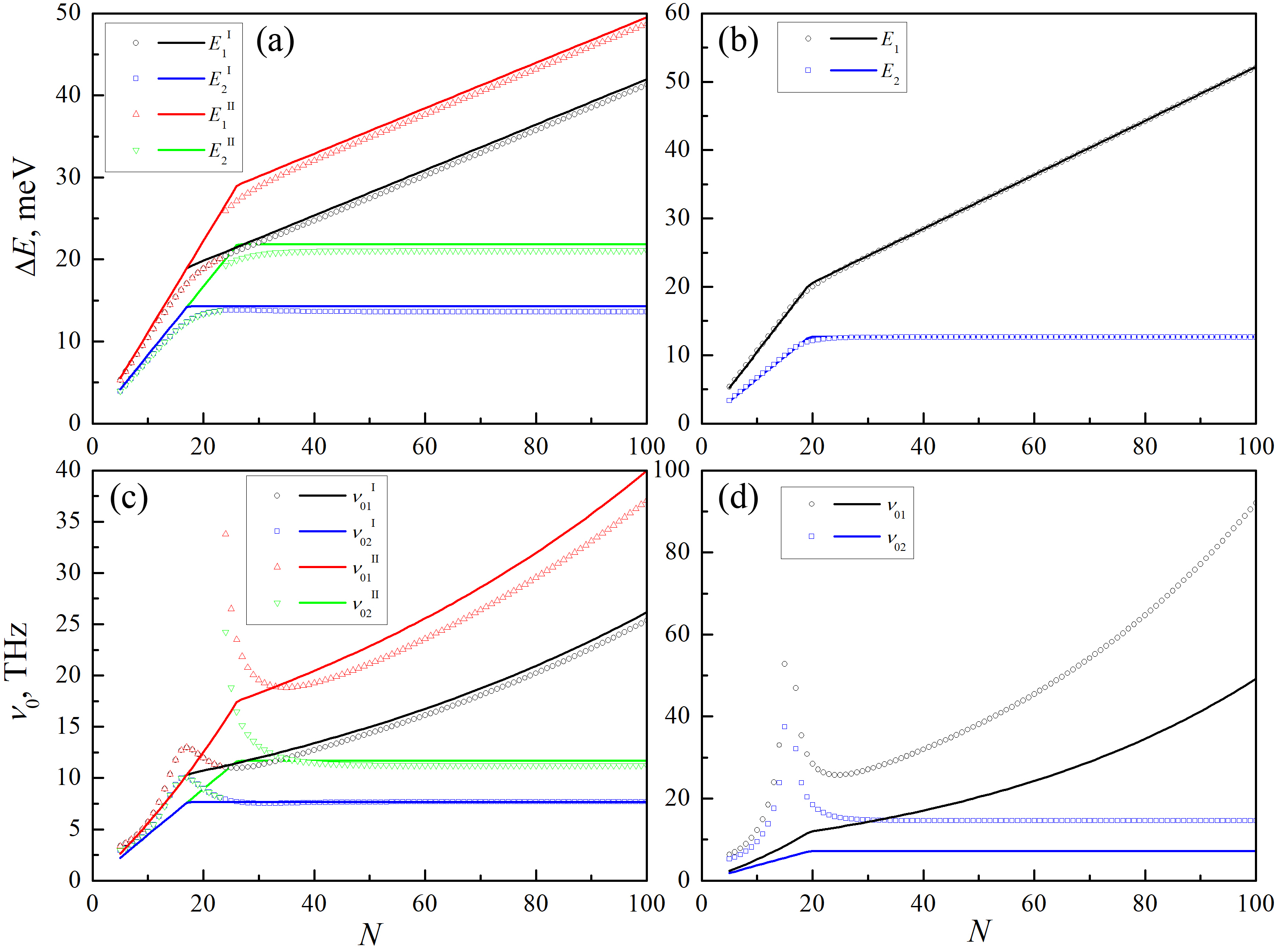}
\caption{\label{fig5}The dependencies of the energy barriers $\Delta E$ (a,b) and the frequency prefactors $\nu_0$ (c,d) on the length $N$ of Co (a,c) and Fe (b,d) atomic chains.
The external magnetic field $B_y$ is 1~T.
The points are the results of the GNEB (a,b) and TST (c,d) calculations.
The solid lines correspond to the theoretical estimation.}
\end{center}
\end{figure*}

Using the results from GNEB calculations, it is straightforward to compute the energy barriers for magnetization reversal in the external magnetic field $B_y$.
The magnetization reversal of the Co chains is characterized by four distinct barriers, as illustrated in Figure~\ref{fig2}(b):
\begin{itemize}
\item the energy barrier ($\Delta E_1^{I}=E_\text{SP1}-E_\text{GS}$) for transition from GS to LM via the formation of ACDW,
\item the energy barrier ($\Delta E_2^{I}=E_\text{SP1}-E_\text{LM}$) for transition from LM to GS via the formation of ACDW,
\item the energy barrier ($\Delta E_1^{II}=E_\text{SP2}-E_\text{GS}$) for transition from GS to LM via the formation of CDW,
\item the energy barrier ($\Delta E_2^{II}=E_\text{SP2}-E_\text{LM}$) for transition from LM to GS via the formation of CDW.
\end{itemize}
Figure~\ref{fig5}(a) illustrates the dependence of energy barriers on the length of the Co chain ($N\in[5,100]$) in an external magnetic field of $B_y=1$~T.
The data points represent the results from numerical GNEB calculations.
As mentioned above, transitions between the GS and LM states can occur through various pathways.
The behavior of short chains resembles that of superparamagnetic particles.
The magnetization reversal takes place without the formation of domain walls.
However, when the Co chain is sufficiently long ($N\ge N_\text{ACDW}=17$ atoms), the transition via the formation of ACDW becomes energetically favorable.
For even longer chains ($N\ge N_\text{CDW}=24$ atoms), a transition involving the formation of CDW also becomes feasible.
It can be observed that the dependencies $\Delta E_1^{I,II}(N)$ and $\Delta E_2^{I,II}(N)$ exhibit different slopes for short chains.
In the case of long chains, the values of $\Delta E_1^{I}(N)$ and $\Delta E_1^{II}(N)$ increase linearly with $N$, while the values of $\Delta E_2^{I}(N)$ and $\Delta E_2^{II}(N)$ tend to constant values.

Figure~\ref{fig5}(b) presents similar dependencies of energy barriers on the length of the Fe chain.
The barriers $\Delta E_1^{I}$ and $\Delta E_1^{II}$ are very close in value, allowing us to neglect the difference between them for the purposes of our analysis.
We will denote the energy barriers for the transition from the GS to the LM state simply as $\Delta E_1$.
Likewise, we can approximate $\Delta E_2^{I}\approx\Delta E_2^{II}\equiv\Delta E_2$.
Similar to the case of the Co chain, there are two mechanisms for the transitions between the GS and LM states in Fe chains.
For shorter Fe chains, magnetization reversal occurs without the formation of domain walls.
However, when the Fe chain exceeds $N_\text{DW}=16$ atoms in length, the transition involving domain wall formation becomes more energetically favorable.
The dependencies $\Delta E_1(N)$ and $\Delta E_2(N)$ exhibit different slopes for short chains.
For long Fe chains, the values of $\Delta E_1(N)$ increase linearly with $N$, while the values of $\Delta E_2(N)$ tend to a constant value.

The frequency prefactors $\nu_0$ calculated within the framework of the harmonic approximation of TST are found to be on the order of $10^{12}-10^{13}$~Hz.
These values are significantly higher than the value of $10^9$~Hz~\cite{Nature416.301}, which is often used as an estimation of the frequency prefactor.
Furthermore, the values of $\nu_0$ also exhibit dependence on the length of the atomic chain $N$, as illustrated in Figures~\ref{fig5}(c) and~\ref{fig5}(d).
For the Co chain, there are four distinct frequency prefactors ($\nu_{01}^{I}$, $\nu_{02}^{I}$, $\nu_{01}^{II}$, and $\nu_{02}^{II}$), each associated with various energy barriers.
At the same time, for the Fe chain, only two different prefactors (($\nu_{01}$ and $\nu_{02}$) are considered.
The naming convention for the frequency prefactors corresponds to the energy barriers they are associated with.
Several interesting features of the dependencies of the frequency prefactors can be highlighted.
(i) For very short atomic chains, the frequency prefactors exhibit a linear increase with the increase of $N$.
The slopes of these dependencies differ between the transition from the GS to the LM state and the reverse transition.
(ii) The frequency prefactors associated with the transition from the LM state to the GS ($\nu_{02}^{I}$ and $\nu_{02}^{II}$ for Co, and $\nu_{02}$ for Fe) tend to constant values as $N$ increases.
(iii) In contrast, the frequency prefactors for the transition from the GS to the LM state ($\nu_{01}^{I}$, $\nu_{01}^{II}$, $\nu_{01}$) exhibit exponential growth with increasing $N$.
(iv) Additionally, the dependencies $\nu_0(N)$ display sharp peaks when the length $N$ of the chain approaches characteristic lengths $N_\text{ACDW}$, $N_\text{CDW}$, or $N_\text{DW}$.
The origin of these peaks is closely related to changes in the mechanism of magnetization reversal.
At specific lengths $N$, the energy landscape undergoes significant alterations, resulting in the emergence of quasi-flat directions characterized by very low eigenvalues $\xi_{\rm sp}^{k}$ (where $k\le2N-1$).
According to Eq. (\ref{eq:prefactor}), this change in the energy landscape leads to a drastic increase in the frequency prefactor.
However, it is important to note that the harmonic approximation of TST becomes invalid in these scenarios.
Consequently, the values at the peaks should be interpreted as indicative of a general trend rather than precise values.
To obtain accurate values for the frequency prefactors in these cases, more sophisticated approximations and calculations would be necessary.
Such detailed calculations are beyond the scope of our current paper.

Let us briefly discuss the influence of DDI on the Co/Pt(332) and Fe/Pt(332) systems.
When comparing the ground states and saddle points with and without DDI, we find that they exhibit almost identical characteristics.
Similar to the observations made in the absence of an external magnetic field~\cite{KOLESNIKOV2023170869,Kolesnikov_EPJB2023}, the energy barriers calculated without considering the DDI are found to be slightly higher than those calculated with the DDI.
However, the difference in these energy barriers is less than 0.1 meV, which is not significant for temperatures above 1 K.
A similar situation arises for the frequency prefactors, where the impact of the DDI is also minimal.
Thus, we can neglect the DDI in our subsequent analytical estimations, simplifying our calculations without greatly compromising accuracy.

\subsection{Analytical estimation of the energy barriers}\label{sec:analytical_barriers}

According to our GNEB results presented in Fig.~\ref{fig3}, we can assume that the magnetization reversal of the Co/Pt(332) chains occurs via the rotation of the magnetic moments in the $XY$-plane.
This assumption substantially simplifies our analytical estimations.
Thus, we begin our discussion by focusing on the Co/Pt(332) system.
The projections of the unit vector ${\bf s}_i$ corresponding to the magnetic moment of the $i$-th atom can be expressed as:
$s_i^x=\sin\theta_i$, $s_i^y=\cos\theta_i$, $s_i^z=0$, where $\theta_i$ is the angle between the unit vector ${\bf s}_i$ and the easy magnetization axis, which is chosen as aligned with the $y$-direction (see Fig.~\ref{fig1}(c)).
We consider that the external magnetic field ${\bf B}$ is applied along the easy magnetization axis, such that $B_x=B_z=0$ and $B_y\ne0$.
Based on our numerical findings, we can also neglect the contributions from DDI in this scenario.
Consequently, the magnetic energy of the Co chain can be written as:
\begin{multline}\label{eq:XYenergy}
E=-J\sum_{i}\cos(\theta_{i+1}-\theta_i)+K_1\sum_{i}\sin^2\theta_i+ \\
+D_z\sum_{i}\sin(\theta_{i+1}-\theta_i)-\mu B_y\sum_{i}\cos\theta_i,
\end{multline}
where $K_1=K-|E|>0$.

To further simplify our analysis, let us assume that the angles $\theta_i$ change only slightly over interatomic distances.
This allows us to treat the discrete angles $\theta_i$ as a continuous function of position, denoting it as $\theta(x)$.
In this framework, we can express the magnetic energy of the infinite Co chain as a functional of $\theta(x)$.
\begin{multline}\label{eq:XYenergy1}
E[\theta]=\frac{1}{a}\int\limits_{-\infty}^{+\infty}\left[\frac{Ja^2}{2}\left(\frac{{\rm d}\theta}{{\rm d}x}\right)^2+
D_za\frac{{\rm d}\theta}{{\rm d}x}+\right.  \\
\left.+K_1\sin^2\theta+\mu B_y(1-\cos\theta)\right]{\rm d}x,
\end{multline}
where $a$ is the nearest-neighbor distance within the atomic chain.
Varying the functional (\ref{eq:XYenergy1}) yields the equation
\begin{equation}\label{eq:XYsinG}
a^2\frac{{\rm d}^2\theta}{{\rm d}x^2}=\frac{2K_1}{J}\sin\theta\left(\cos\theta+\chi\right),
\end{equation}
where $\chi=\mu B_y/2K_1$.

The equation (\ref{eq:XYsinG}) has a trivial solution given by $\theta(x)=0$, which corresponds to an energy $E=0$.
This solution represents the ground state of the infinitely long atomic chain, where all magnetic moments are aligned with the external magnetic field ($B_y>0$).
Next, we aim to find the low-energy excited state of the chain as a solution to Eq.~(\ref{eq:XYsinG}) with finite energy ($E<\infty$).
It is important to note that the solutions $\theta(x)=\pi$ or those taking the form of single domain walls~\cite{Landau_book8} do not satisfy the condition of $E<\infty$.
Therefore, we should find a solution to Eq.~(\ref{eq:XYsinG}) under the boundary conditions $\theta(x)\to0$ as $x\to\pm\infty$.
Employing standard methods~\cite{Landau_book8, Rajaraman_book} leads us to solutions in the form of twisted or untwisted domain wall pairs~\cite{PhysRevB.50.16485}.
For our purposes, it is convenient to express these solutions in the following form
\begin{multline}\label{eq:asymmitric}
\Theta(x,K_1,B_y)=\\
=\arccos\frac{f(\chi)^2-2f(\chi)(2+3\chi)\phi(x)+\chi^2\phi(x)^2}{f(\chi)^2+2f(\chi)(2+\chi)\phi(x)+\chi^2\phi(x)^2},
\end{multline}
where
\begin{equation}\label{eq:phi}
\phi(x)=\exp\left(\sqrt{\frac{8K_1}{J}(1+\chi)}\frac{x-x_0}{a}\right)
\end{equation}
and $f(\chi)=2+3\chi+2\sqrt{(1+\chi)(1+2\chi)}$.
The solution (\ref{eq:asymmitric}) with $\chi>0$ corresponds to a twisted domain wall pair, while the solution with $\chi<0$ represents an untwisted domain wall pair.
In the limit as $\chi\to0$, the solution (\ref{eq:asymmitric}) approaches a superposition of two domain walls.
Thus, we conclude that the solution (\ref{eq:asymmitric}) indeed describes a coupled state of two domain walls in the presence of the external magnetic field.
One can see that the solution (\ref{eq:asymmitric}) yields $\Theta(x_0,\chi)=\pi/2$ at some arbitrary point $x_0$.
By substituting this solution into the functional (\ref{eq:XYenergy1}), we can determine the energy of this state~\cite{PhysRevLett.65.787}
\begin{equation}\label{eq:Estate}
E=\sqrt{8K_1J}\left(2\sqrt{1+\chi}+\chi\ln\frac{1+\sqrt{1+\chi}}{1-\sqrt{1+\chi}}\right).
\end{equation}
The first term in (\ref{eq:Estate}) represents the energy of two non-interacting domain walls, while the second term accounts for the energy of the magnetic moments situated between these domain walls.

Now, let us consider the semi-infinite atomic chain defined for $x\in(-\infty,0]$.
Within the framework of the continuous model, the magnetic energy is analogous to the functional presented in (\ref{eq:XYenergy1}), but with the upper limit of the integral set to 0 instead of $\infty$.
Varying the energy functional once again leads to the equation (\ref{eq:XYsinG}).
The function (\ref{eq:asymmitric}) remains a valid solution, satisfying the boundary condition $\theta(x)\to0$ as $x\to-\infty$.
By substituting this solution into the energy functional, we obtain the energy of the semi-infinite atomic chain as a function $E=E(\theta_0)$.
Here, $\theta_0=\theta(0)$ represents the angle between the unit vector ${\bf e}_y$ and the direction of the magnetic moment at the end of the chain.
The extrema of the function $E(\theta_0)$ can be determined by solving the equation ${\rm d}E/{\rm d}\theta_0=0$.

After performing the straightforward calculations, we find that the function $E(\theta_0)$ has an extremum
\begin{equation}\label{eq:XYenergy4}
E_{\rm end}(K_1,D_z,B_y)\approx-\frac{D_z^2}{\sqrt{8K_1J(1+\chi)}}
\end{equation}
 at the angle
\begin{equation}\label{eq:XYtheta0}
\theta_0\approx-\frac{D_z}{\sqrt{2K_1J(1+\chi)}}.
\end{equation}
The value given in (\ref{eq:XYenergy4}) represents the decrease in the energy of the atomic chain resulting from the anticlockwise rotation of the magnetic moments at the free end by the angle specified in (\ref{eq:XYtheta0}).
If $\chi=0$, the Eqs. (\ref{eq:XYenergy4}) and (\ref{eq:XYtheta0}) reduce to those presented in Ref.~\cite{KOLESNIKOV2023170869}.
The function $E(\theta_0)$ has another extremum
\begin{multline}\label{eq:Estate1}
E_{\rm DW}(K_1,D_z,B_y)=\sqrt{8K_1J}\left(\sqrt{1+\chi}+\right.\\
\left.+\frac{D_z}{\sqrt{8K_1J}}\arccos(-2\chi-1)
+\frac{\chi}{2}\ln\frac{1+\sqrt{1+\chi}}{1-\sqrt{1+\chi}}\right),
\end{multline}
which corresponds to the solution given in (\ref{eq:asymmitric}) at $x_0=0$.
The value given in (\ref{eq:Estate1}) represents the energy of the domain wall situated closer to the right end of the semi-infinite atomic chain.
When $\chi=0$, the domain wall shifts away from the end of the chain.
The energy $E_{\rm DW}(K_1,\pm D_z,B_y)$ transforms to the energy of CDW ($\sqrt{8K_1J}+\pi D_z$) or ACDW ($\sqrt{8K_1J}-\pi D_z$) positioned in the middle of the chain~\cite{KOLESNIKOV2023170869,PRB.78.140403}.

The solution given in (\ref{eq:asymmitric}) can be utilized to approximate the numerical results shown in Fig.~~\ref{fig3}.
We assume that $B_y>0$.
The ground state, local minimum, and saddle points can be approximated as
\begin{multline}\label{eq:groundstate}
\theta_i^{\rm GS}(i_1)=\Theta\left((i_0-i_1-i)a,K_1,B_y\right)-\\
-\Theta\left((i-i_0-i_1)a,K_1,B_y\right),
\end{multline}
\begin{multline}\label{eq:localminimum}
\theta_i^{\rm LM}(i_2)=\Theta\left((i_0-i_2-i)a,K_1,-B_y\right)-\\
-\Theta\left((i-i_0-i_2)a,K_1,-B_y\right)+\pi,
\end{multline}
\begin{multline}\label{eq:saddlepoint1}
\theta_i^{\rm SP1}(i_3,i_4)=\Theta\left((i_3-i)a,K_1,-B_y\right)-\\
-\Theta\left((i-i_0-i_4)a,K_1,-B_y\right)+\pi,
\end{multline}
\begin{multline}\label{eq:saddlepoint2}
\theta_i^{\rm SP2}(i_5,i_6,i_7)=\Theta\left((i_0-i_5-i)a,K_1,-B_y\right)+\\
+\Theta\left((i-i_6)a,K_1,-B_y\right)-\Theta\left((i-i_0-i_7)a,K_1,B_y\right)+\pi,
\end{multline}
where $i_0=(N+1)/2$ is the midpoint of the chain, and $i_k$ (for $k=\overline{1,7}$) are the fitting parameters.

Substituting the approximations (\ref{eq:groundstate}), (\ref{eq:localminimum}), (\ref{eq:saddlepoint1}), and (\ref{eq:saddlepoint2}) into Eq.~(\ref{eq:XYenergy}) yields the energies of the GS, LM, SP1, and SP2 states as functions of the parameters $i_k$: $E_{\rm GS}(i_1)$, $E_{\rm LM}(i_2)$, $E_{\rm SP1}(i_3,i_4)$, and $E_{\rm SP2}(i_5,i_6,i_7)$.
To obtain the optimal values of $\tilde i_k$ (for $k=\overline{1,7}$), we minimize the functions by adjusting the parameters $i_1$, $i_2$, $i_4$, $i_5$, $i_7$ and maximize them with respect to the parameters $i_3$, $i_6$.
Subsequently, we determine the optimal values for $\tilde s_{i,x}=\sin\tilde\theta_i$, $\tilde s_{i,y}=\cos\tilde\theta_i$, and $\tilde s_{i,z}=0$, where
$\tilde\theta_i^{\rm GS}=\theta_i^{\rm GS}(\tilde i_1)$,
$\tilde\theta_i^{\rm LM}=\theta_i^{\rm LM}(\tilde i_2)$,
$\tilde\theta_i^{\rm SP1}=\theta_i^{\rm SP1}(\tilde i_3,\tilde i_4)$, and
$\tilde\theta_i^{\rm SP2}=\theta_i^{\rm SP2}(\tilde i_5,\tilde i_6,\tilde i_7)$.
The magnetic configurations $\tilde s_{i,x}$, $\tilde s_{i,y}$, $\tilde s_{i,z}$ of the Co chain, consisting of $N=100$ atoms in the GS, LM, SP1, and SP2 states, are shown in Fig.~\ref{fig3} with solid lines.
There is a remarkable agreement between the values of $s_{i,x}$, $s_{i,y}$ and $\tilde s_{i,x}$, $\tilde s_{i,y}$, respectively.
This result indicates that the magnetization reversal of the Co chains can be effectively described within the framework of the $XY$-model, similar to the case of zero external magnetic field~\cite{KOLESNIKOV2023170869}.
It is intriguing to compare the values of the angle $\theta_0$ across the different approximations discussed above.
According to the GNEB calculations, $\theta_0=6.88^\circ$ in the ground state (with $N=100$ and $B_y=1$~T).
The discrete approximation given by (\ref{eq:groundstate}) yields $\tilde\theta_0=6.85^\circ$.
The continuous approximation from (\ref{eq:XYtheta0}) results in a value of $7.33^\circ$.
In the case of the LM state, we find $\theta_0=7.47^\circ$ and $\tilde\theta_0=7.43^\circ$.
The continuous approximation (\ref{eq:XYtheta0}) provides a value of $7.87^\circ$.
It is not surprising that the estimates derived from the continuous model are somewhat less accurate than those from the discrete $XY$-model.
However, the difference is minimal ($\Delta\theta<0.5^\circ$).
Therefore, we can use the results from the continuous model for estimating the energy barriers.

Let us assume that the collinear state $\theta_i=0$ (for $i=\overline{1,N}$) has zero energy.
In this case, the energy of the collinear state $\theta_i=\pi$ is $2\mu B_y N$.
The approximations given by (\ref{eq:groundstate}) and (\ref{eq:localminimum}) indicate that the GS and LM states differ from the collinear states only by deviations at the ends of the chain.
Therefore, their energies can be estimated within the framework of the continuous model as $E_{\rm GS}=2E_{\rm end}(K_1,D_z,B_y)$ and $E_{\rm LM}=2\mu B_y N+2E_{\rm end}(K_1,D_z,-B_y)$.
Furthermore, the approximations (\ref{eq:saddlepoint1}) and (\ref{eq:saddlepoint2}) suggest that the energies of the SP1 and SP2 states can be estimated as follows: $E_{\rm SP1}=2\mu B_y N+E_{\rm end}(K_1,D_z,-B_y)+E_{\rm DW}(K_1,-D_z,-B_y)$ and
$E_{\rm SP2}=2\mu B_y N+E_{\rm end}(K_1,D_z,-B_y)+E_{\rm DW}(K_1,D_z,-B_y)$.
Thus, the energy barriers for the magnetization reversal of the long Co chain can be estimated as
\begin{multline}\label{eq:EdI1}
\Delta E^{I,{\rm long}}_1=E_{\rm SP1}-E_{\rm GS}=2\mu B_y N+\\
+E_{\rm DW}(K_1,-D_z,-B_y)+E_{\rm end}(K_1,D_z,-B_y)-\\
-2E_{\rm end}(K_1,D_z,B_y),
\end{multline}
\begin{multline}\label{eq:EdI2}
\Delta E^{I,{\rm long}}_2=E_{\rm SP1}-E_{\rm LM}=E_{\rm DW}(K_1,-D_z,-B_y)-\\
-E_{\rm end}(K_1,D_z,-B_y),
\end{multline}
\begin{multline}\label{eq:EdII1}
\Delta E^{II,{\rm long}}_1=E_{\rm SP2}-E_{\rm GS}=2\mu B_y N+\\
+E_{\rm DW}(K_1,D_z,-B_y)+E_{\rm end}(K_1,D_z,-B_y)-\\
-2E_{\rm end}(K_1,D_z,B_y),
\end{multline}
\begin{multline}\label{eq:EdII2}
\Delta E^{II,{\rm long}}_2=E_{\rm SP2}-E_{\rm LM}=E_{\rm DW}(K_1,D_z,-B_y)-\\
-E_{\rm end}(K_1,D_z,-B_y).
\end{multline}

Now, let us consider short Co chains.
In this scenario, we can assume that all the angles $\theta_i$ are identical ($\theta_i=\theta$).
Consequently, the magnetic energy of a chain consisting of $N$ atoms can be expressed as
\begin{equation}\label{eq:Esimple}
E^{\rm short}=NK_1(\sin^2\theta+2\chi(1-\cos\theta)).
\end{equation}
It is assumed that the energy of the ground state ($\theta=0$) is zero.
The extrema of this energy can be determined from the equation ${\rm d}E^{\rm short}/{{\rm d}\theta}=0$.
If $\chi\in[0,1]$, then the energy exhibits two local minima: $E=0$ at $\theta=0$ and $E=4NK_1\chi$ at $\theta=\pi$, along with a local maximum
\begin{equation}\label{eq:ED1simple}
\Delta E_1^{\rm short}=NK_1(1+\chi)^2
\end{equation}
at $\cos\theta=-\chi$.
Eq.~(\ref{eq:ED1simple}) represents the energy barrier for the transition from the GS to the LM state.
The energy barrier for the transition from the LM state to the GS is the following
\begin{equation}\label{eq:ED2simple}
\Delta E_2^{\rm short}=NK_1(1-\chi)^2.
\end{equation}
The results for both long and short chains can be combined as follows:
$\Delta E^{I}_1=\min(\Delta E_1^{\rm short},\Delta E^{I,{\rm long}}_1)$,
$\Delta E^{I}_2=\min(\Delta E_2^{\rm short},\Delta E^{I,{\rm long}}_2)$,
$\Delta E^{II}_1=\min(\Delta E_1^{\rm short},\Delta E^{II,{\rm long}}_1)$,
$\Delta E^{II}_2=\min(\Delta E_2^{\rm short},\Delta E^{II,{\rm long}}_2)$.
Utilizing this approximation, we can estimate the energy barriers for any length $N$ of the Co atomic chain.
The dependencies $\Delta E(N)$ in the external magnetic field $B_y=1$~T are presented in Fig.~\ref{fig5}(a) with solid lines.
It is evident that the theoretical estimates align well with the numerical GNEB results, except in the range where the magnetization reversal mechanism undergoes a change.

Now, let us discuss the Fe chains on the Pt(332) surface.
For short atomic chains, the Eqs. (\ref{eq:ED1simple}) and  (\ref{eq:ED2simple}) remain applicable.
In the case of long atomic chains, all magnetic moments are oriented within the $XY$-plane in both the GS and LM states.
Therefore, the GS and LM states can be approximated using the functions (\ref{eq:groundstate}) and (\ref{eq:localminimum}), where $K_1=K-|E|$ is replaced by $K_2=K+|E|$:
\begin{multline}\label{eq:groundstateFe}
\theta_i^{\rm GS}(i_1)=\Theta\left((i_0-i_1-i)a,K_2,B_y\right)-\\
-\Theta\left((i-i_0-i_1)a,K_2,B_y\right),
\end{multline}
\begin{multline}\label{eq:localminimumFe}
\theta_i^{\rm LM}(i_2)=\Theta\left((i_0-i_2-i)a,K_2,-B_y\right)-\\
-\Theta\left((i-i_0-i_2)a,K_2,-B_y\right)+\pi.
\end{multline}
The optimal values $\tilde i_1$ and $\tilde i_2$ can be determined by minimizing the magnetic energy of the chain.
Subsequently, the values of $\tilde s_{i,x}$ and $\tilde s_{i,y}$ can be calculated.
These values for the Fe chain consisting of $N=100$ atoms in both GS and LM states are depicted in Fig.~\ref{fig4} with solid lines.
Similar to the Co chain, there is a remarkable agreement between the values of $s_{i,x}$, $s_{i,y}$ and $\tilde s_{i,x}$, $\tilde s_{i,y}$, respectively.
Therefore, the continuous approximations (\ref{eq:XYenergy4}) and (\ref{eq:XYtheta0}) are also applicable to the Fe chains, with $K_1$ being replaced by $K_2$.
Let us compare the values of the angle $\theta_0$ obtained from different approximations.
According to the GNEB calculations, $\theta_0=5.94^\circ$ in the ground state ($N=100$, $B_y=1$~T).
The discrete approximation (\ref{eq:groundstateFe}) yields $\tilde\theta_0=5.93^\circ$, while the continuous approximation provides a value of $6.92^\circ$.
For the LM state, we find $\theta_0=6.22^\circ$ and $\tilde\theta_0=6.21^\circ$.
The continuous approximation results in a value of $7.18^\circ$.
As was observed with the Co chain, the estimates derived from the continuous model are less accurate than those obtained from the discrete $XY$-model.
However, the differences are again minimal.

The situation becomes more intriguing when the domain wall is positioned in the middle of the Fe atomic chain.
On one hand, magnetic anisotropy tends to orient the domain wall into the $YZ$-plane.
On the other hand, the DMI encourages a rotation of the domain wall into the $XY$-plane.
Consequently, the domain walls in the Fe chains exhibit an intermediate structure that lies between Bloch and N\'eel walls.
Following Ref.~\cite{Kolesnikov_EPJB2023}, we assume that all the magnetic moments within the domain wall lie in the$X^\prime Y$-plane (see Fig.~\ref{fig1}(c)).
This assumption implies that all angles $\phi_i$ are uniform, such that $\phi_i=\phi_0$.
The projections of the unit vector ${\bf s}_i$ can then be expressed as:
$s_i^{x}=s_i^{\prime x}\cos\varphi_0=\sin\theta_i\cos\varphi_0$,
$s_i^{y}=s_i^{\prime y}=\cos\theta_i$,
$s_i^{z}=-s_i^{\prime x}\sin\varphi_0=-\sin\theta_i\sin\varphi_0$.
By substituting these expressions into the Hamiltonian, we obtain the energy equations analogous to (\ref{eq:XYenergy}) and (\ref{eq:XYenergy1}), where the parameters $K_1$ and $D_z$ are modified to $K_3=K+|E|\cos(2\varphi_0)$ and $\tilde D=D_z\cos\varphi_0$, respectively.
Equations (\ref{eq:XYsinG}), (\ref{eq:asymmitric}), (\ref{eq:phi}), (\ref{eq:Estate}), and (\ref{eq:Estate1}) should be adjusted accordingly.
It is noteworthy that the terms in Eqs. (\ref{eq:XYenergy}) and (\ref{eq:XYenergy1}) that depend on the angle $\phi_0$ do not rely on the external magnetic field.
Therefore, the value of $\phi_0$ remains unaffected by external magnetic fields.
It takes the same value as in the absence of an external magnetic field~\cite{Kolesnikov_EPJB2023}:
\begin{equation}\label{eq:phy0}
\varphi_0=\pm\arccos\sqrt{\frac{K-|E|}{2|E|\left(\frac{16J|E|}{\pi^2D_z^2}-1\right)}}.
\end{equation}
The plus sign corresponds to the SP1 state, while the minus sign corresponds to the SP2 state.
We can now compare the numerical and analytical estimations of the angle $\phi_0$.
For the states shown in Fig.~\ref{fig4}, numerical calculations yield values of $81.41^\circ$ for the SP1 state and $-81.33^\circ$ for the SP2 state.
The analytic expression in Eq.~(\ref{eq:phy0}) gives $\phi_0=\pm 80.43^\circ$.
One can see that there is a good agreement between these values.

Now, let us discuss the approximations of the SP1 and SP2 states presented in Fig.~\ref{fig4}.
For the approximation of the SP1 state, we utilize the function
\begin{equation}\label{eq:saddlepointFe-1}
\theta_{1i}^{\rm SP}(i_3)=\Theta\left((i-i_3)a,K_3,-B_y\right),
\end{equation}
which models the domain wall located in the middle of the chain, along with the function
\begin{multline}\label{eq:saddlepointFe-2}
\theta_{2i}^{\rm SP}(i_4,i_5)=\Theta\left((i-i_4)a,K_2,B_y\right)+\\
+\Theta\left((i_5-i)a,K_2,-B_y\right)-\pi,
\end{multline}
which accounts for the deviations at the ends of the chain.
The projections of the unit vector ${\bf s}_i$ can be approximated as follows:
\begin{equation}\label{eq:xSPFe}
s^x_i=\left(\sin\theta_{1i}^{\rm SP}\cos\varphi_0+\sin\theta_{2i}^{\rm SP}\right)/s_i,
\end{equation}
\begin{equation}\label{eq:ySPFe}
s^y_i=\cos\left(\theta_{1i}^{\rm SP}+\theta_{2i}^{\rm SP}\right)/s_i,
\end{equation}
\begin{equation}\label{eq:zSPFe}
s^z_i=\sin\theta_{1i}^{\rm SP}\sin\varphi_0/s_i,
\end{equation}
where
\begin{equation}\label{eq:absSPFe}
s_i=\sqrt{(s^x_i)^2+(s^y_i)^2+(s^z_i)^2}.
\end{equation}
For the approximation of the SP2 state, we need to replace $s^z_i$ with $-s^z_i$.
Minimizing the magnetic energy with respect to the parameters $i_4$ and $i_5$, while maximizing with respect to the parameter $i_3$, yields the optimal values $\tilde s^x_i$, $\tilde s^y_i$, and $\tilde s^z_i$, which are presented in Fig.~\ref{fig4} as solid lines.
It is important to note that the approximation of the saddle points for the Fe chains is not as accurate as for the Co chains.
Nevertheless, we observe a qualitative agreement between the numerical results and the theoretical approximations.
Therefore, we can utilize the continuous model for estimating the energy barriers.

Within the framework of the continuous model, the energies of the GS and the LM states can be estimated as $E_{\rm GS}=2E_{\rm end}(K_2,D_z,B_y)$ and $E_{\rm LM}=2\mu B_y N+2E_{\rm end}(K_2,D_z,-B_y)$, respectively.
According to the approximations given by equations (\ref{eq:saddlepointFe-1}) and (\ref{eq:saddlepointFe-2}), the energy of the saddle points can be estimated as $E_{\rm SP}=2\mu B_y N+E_{\rm DW}(K_3,-\tilde D,-B_y)+E_{\rm end}(K_2,D_z,B_y)+E_{\rm end}(K_2,D_z,-B_y)$.
Consequently, the energy barriers for the magnetization reversal of long Fe chains can be calculated as
\begin{multline}\label{eq:Ed1Fe}
\Delta E^{\rm long}_1=E_{\rm SP}-E_{\rm GS}=2\mu B_y N+\\
+E_{\rm DW}(K_3,-\tilde D,-B_y)+E_{\rm end}(K_2,D_z,-B_y)-\\
-E_{\rm end}(K_2,D_z,B_y),
\end{multline}
\begin{multline}\label{eq:Ed2Fe}
\Delta E^{\rm long}_2=E_{\rm SP}-E_{\rm LM}=E_{\rm DW}(K_3,-\tilde D,-B_y)+\\
+E_{\rm end}(K_2,D_z,B_y)-E_{\rm end}(K_2,D_z,-B_y).
\end{multline}
Finally, the results for both long and short chains can be summarized as
$\Delta E_1=\min(\Delta E_1^{\rm short},\Delta E^{\rm long}_1)$ and $\Delta E_2=\min(\Delta E_2^{\rm short},\Delta E^{\rm long}_2)$.
This approximation allows us to estimate the energy barriers for any length $N$ of the Fe atomic chain.
The dependencies $\Delta E(N)$ in the presence of an external magnetic field $B_y=1$~T are shown in Fig.~\ref{fig5}(b) with solid lines.
One can see that the theoretical estimates are in the excellent agreement with the numerical results from the GNEB method, except in the region where the mechanism of magnetization reversal changes.
A comparison of Figs.~\ref{fig5}(a) and \ref{fig5}(b) reveals that the analytical estimates of the energy barriers for the Fe chains are even more accurate than those for the Co chains.

\subsection{Analytical estimation of the frequency prefactors}\label{sec:analytical_prefactors}

The calculation of the frequency prefactor $\nu_0$ in the harmonic approximation of TST requires finding the eigenvalues $\xi^{k}$ of the $2N$-dimensional Hessian matrix $\mathcal{H}$ of energy at the local minima and saddle points (see Eq. (\ref{eq:prefactor})).
An analytical solution to the eigenvalue problem is feasible only in certain simple cases for infinitely long chains~\cite{PhysRev.130.1677,PhysRevB.50.16485,PhysRevB.50.16501}.
For finite-length atomic chains, the magnetic moments at the ends and in the middle of the chain are in nonequivalent conditions, which complicates the eigenvalue problem significantly.
A general algorithm for solving the eigenvalue problem can be implemented numerically.
The numerical results for $\nu_0$ are presented in Fig.~\ref{fig5}.
However, it is valuable to derive a simple theoretical estimation of the frequency prefactor, because this can provide insight into how $\nu_0$ depends on the model parameters.
According to Eq.~(\ref{eq:frequency}), the magnetization reversal frequency is an exponential function of the energy barrier and a linear function of the frequency prefactor.
Thus, a highly accurate estimation of the frequency prefactor is not necessary.
In fact, the theoretical estimate of the frequency prefactor should only satisfy the following two conditions:
(i) it should be correct in order of magnitude, and
(ii) it should reproduce the fundamental patterns of dependence on the chain length.

Let us begin with the simplest case, where one magnetic moment is subjected to an external magnetic field ${\bf B}=B_y{\bf e}_y$.
The energy of the magnetic moment equals to
\begin{equation}\label{eq:Energy1atom}
E_1=-K(s^y)^2+E\left((s^z)^2-(s^x)^2\right)-\mu B_y s^y.
\end{equation}
For definiteness, we will assume that $E>0$ (as is the case for Co chains) and $B_y>0$.
It is easy to find that the ground state is given by ${\bf s}_{\rm GS}=(0,1,0)$, the local minimum by ${\bf s}_{\rm LM}=(0,-1,0)$, and the saddle point by ${\bf s}_{\rm SP}=(\eta,\xi,0)$, with and $\xi=-\mu B_y/2(K-E)$ and $\eta=\sqrt{1-\xi^2}$).
The eigenvalues and corresponding eigenvectors of the Hessian matrix in the ground state are $\xi^1_{\rm GS}=2(K+E)+\mu B_y$, ${\bf e}^1_{\rm GS}=(0,0,1)$ and $\xi^2_{\rm GS}=2(K-E)+\mu B_y$, ${\bf e}^2_{\rm GS}=(1,0,0)$.
In the local minimum, the analogous values are $\xi^1_{\rm LM}=2(K+E)-\mu B_y$, ${\bf e}^1_{\rm LM}=(0,0,1)$ and $\xi^2_{\rm LM}=2(K-E)-\mu B_y$, ${\bf e}^2_{\rm LM}=(1,0,0)$.
At the saddle point, we find $\xi^1_{\rm SP}=4E$, ${\bf e}^1_{\rm SP}=(0,0,1)$ and $\xi^2_{\rm SP}=-2(K-E)+\frac{(\mu B_y)^2}{2(K-E)}<0$, ${\bf e}^2_{\rm SP}=(\xi,-\eta,0)$.
Using Eq.~(\ref{eq:prefactor}), we can find the frequency prefactors
\begin{equation}\label{eq:prefactor01}
\nu_{01}^{(1)}=\frac{\gamma}{\pi\mu}\sqrt{\left(K+\frac{\mu B_y}{2}\right)^2-E^2},
\end{equation}
\begin{equation}\label{eq:prefactor02}
\nu_{02}^{(1)}=\frac{\gamma}{\pi\mu}\sqrt{\left(K-\frac{\mu B_y}{2}\right)^2-E^2}.
\end{equation}
The frequency prefactors given by Eqs.~(\ref{eq:prefactor01}) and (\ref{eq:prefactor02}) are independent of the sign of $E$.
Thus, the presented results are applicable for both Co and Fe atoms on the Pt(332) surface.
In the absence of the external magnetic field, the frequency prefactors become identical: $\nu_{0}^{(1)}=\frac{\gamma}{\pi\mu}\sqrt{K^2-E^2}$.
This simple estimation yields $\nu_{0}^{(1)}=0.47$~THz for Co and $\nu_{0}^{(1)}=0.40$~THz for Fe.
These values are of the same order of magnitude as the numerical results presented in Fig.~\ref{fig5}(c,d).

Now, let us consider a chain consisting of $N$ atoms in the absence of an external magnetic field ($B_y=0$).
We assume that $E>0$.
Unfortunately, the eigenvalue problem for a finite-length atomic chain with DMI is too complex for an analytical solution.
Therefore, we make the following simplifications:
(i) we neglect DMI,
(ii) we consider the chain with periodic boundary conditions rather than one with free ends, and
(iii) we focus on short atomic chains where the magnetization reversal occurs without the formation of domain walls.
In this simplified scenario, the directions of all magnetic moments in the GS (LM) and SP states are the same as those in the case of a single magnetic moment.
After the diagonalization of the Hessian matrix in the GS (LM) and SP states, we can compute
\begin{equation}\label{eq:prefactorNatom_B=0}
\nu_{0}^{(N),{\rm short}}=N\frac{\gamma}{\pi\mu}\sqrt{K^2-E^2}\sqrt{F(J,K,E,N)},
\end{equation}
where
\begin{multline}\label{eq:Function}
F(J,K,E,N)=
\frac{\prod\limits_{n=1}^{N-1}\left[K+E+J\left(1-\cos\left(\frac{2\pi n}{N}\right)\right)\right]}
{\prod\limits_{n=1}^{N-1}\left[2E+J\left(1-\cos\left(\frac{2\pi n}{N}\right)\right)\right]}\cdot\\
\cdot\frac{\prod\limits_{n=1}^{N-1}\left[K-E+J\left(1-\cos\left(\frac{2\pi n}{N}\right)\right)\right]}
{\prod\limits_{n=1}^{N-1}\left[E-K+J\left(1-\cos\left(\frac{2\pi n}{N}\right)\right)\right]}.
\end{multline}
The function $F(J,K,E,N)$ must remain positive, imposing a limitation on the model parameters.
The atomic chain must be shorter than a maximal length
\begin{equation}\label{eq:Nmax}
N_{\max}=\frac{2\pi}{\arccos\left(\frac{J-K+E}{J}\right)}.
\end{equation}
Specifically, we find $N_{\max}=35.42$ for Co chains and $N_{\max}=32.78$ for Fe chains.
Figure~\ref{fig5} shows that the mechanism of magnetization reversal changes at $N<N_{\max}$.
This indicates that Eq.~(\ref{eq:prefactorNatom_B=0}) is valid for all short Co and Fe chains.
The function (\ref{eq:Function}) increases monotonically from $F\approx1$ at $N\ll N_{\max}$ to $F\to\infty$ at $N\to N_{\max}$.
For sufficiently short atomic chains, $F\approx1$, leading to the approximation $\nu_{0}^{(N),{\rm short}}\approx N\nu_{0}^{(1)}$.
This simple relationship qualitatively agrees with our numerical results presented in Fig.~\ref{fig5}(c,d).

By comparing the energy barrier for magnetization reversal in a short atomic chain, given by $\Delta E^{\rm short}=N(K-E)$, with the analogous barrier for a single atom, $\Delta E^{(1)}=(K-E)$, we can express the relationship as $N=\Delta E^{\rm short}/\Delta E^{(1)}$.
Now, let us consider the magnetization reversal of a long atomic chain.
The formation of a domain wall at the end of the chain is associated with the reversal of a small number of magnetic moments.
This number can be roughly estimated as $N_{\rm eff}=E_{\rm DW}/\Delta E^{(1)}$.
With the same accuracy, the frequency prefactor can be estimated as $\nu_{0}^{\rm long}\approx N_{\rm eff}\nu_{0}^{(1)}$.
It is important to note that in the absence of an external magnetic field, we have $\Delta E\approx E_{\rm DW}$.
Therefore, we find that $\nu_{0}^{\rm long}\sim N_{\rm eff}\sim \Delta E$, which is consistent with the Meyer-Neldel rule~\cite{Meyer_Neldel}.
This empirical rule has been observed in various fields, including materials science, physics~\cite{PhysRevLett.75.469}, chemistry, and biology.
This lends some justification to our approach.

Now, we can consider the magnetization reversal of the atomic chain in the presence of an external magnetic field.
The transition from the LM state to the ground state is associated with the reversal of a small number of magnetic moments that form the domain wall.
Thus, we can estimate the frequency prefactor in a manner analogous to the case of zero magnetic field:
\begin{equation}\label{eq:prefactor02a}
\nu_{02}=\nu_{02}^{(1)}N_{\rm eff},
\end{equation}
where
\begin{equation}\label{eq:prefactor02CoI}
N_{\rm eff}^{I,{\rm Co}}=\min\left(N,\frac{E_{\rm DW}(K_1,-D_z,-B_y)}{K_1(1-\chi)^2}\right),
\end{equation}
\begin{equation}\label{eq:prefactor02CoII}
N_{\rm eff}^{II,{\rm Co}}=\min\left(N,\frac{E_{\rm DW}(K_1,D_z,-B_y)}{K_1(1-\chi)^2}\right),
\end{equation}
for Co chains, and
\begin{equation}\label{eq:prefactor02Fe}
N_{\rm eff}^{\rm Fe}=\min\left(N,\frac{E_{\rm DW}(K_3,-\tilde D,-B_y)}{K_1(1-\chi)^2}\right),
\end{equation}
for Fe chains.
Theoretical estimates of the frequency prefactors $\nu_{02}$ are presented in Fig.~\ref{fig5}(c,d) as solid lines.
These estimates qualitatively reproduce the key behaviors observed in the numerical results:
(i) a linear increase in the frequency prefactors for very short atomic chains, and
(ii) the tendency to reach constant values for longer atomic chains.
Notably, the theoretical estimates closely match the numerical results for long Co chains.
Indeed, for $N=100$ and $B_y=1$~T, the numerical calculations yield $\nu_{02}^{I}=7.68$~THz and $\nu_{02}^{II}=11.2$~THz, whereas the theoretical estimates provide values of $7.66$~THz and $11.7$~THz, respectively.
In contrast, for Fe chains, the agreement is only in the order of magnitude.
For $N=100$ and $B_y=1$~T, the numerical calculations yield $\nu_{02}=14.7$~THz, which is approximately twice the theoretical estimate of $7.23$~THz.

The theoretical approach to the transition from the GS to the LM state is more complex, as this transition involves the reversal of nearly all magnetic moments.
Consequently, the frequency prefactor must be adjusted by multiplying it by the factor $\sqrt{\det\mathcal{H}[s_{GS}]/\det\mathcal{H}[s_{LM}]}$.
Carrying out straightforward calculations yields the expression
\begin{multline}\label{eq:detH_GS}
\det\mathcal{H}[s_{GS}]=\\
=\prod\limits_{n=1}^{N}2\left[K+E+\frac{\mu B_y}{2}+J\left(1-\cos\left(\frac{2\pi n}{N}\right)\right)\right]\\
\prod\limits_{n=1}^{N}2\left[K-E+\frac{\mu B_y}{2}+J\left(1-\cos\left(\frac{2\pi n}{N}\right)\right)\right].
\end{multline}
To determine $\det\mathcal{H}[s_{LM}]$, we need to replace $B_y\to-B_y$ in Eq.~(\ref{eq:detH_GS}).
Let us introduce the function $\psi(E,B_y)=(K+E+\frac{\mu B_y}{2})/J$.
Assuming $N\gg1$, the products in Eq.~(\ref{eq:detH_GS}) can be easily calculate.
The first product is equal to
\begin{multline}\label{eq:detH_GS1}
\prod\limits_{n=1}^{N}2J\left[1+\psi(E,B_y)-\cos\left(\frac{2\pi n}{N}\right)\right]=\\
=(2J)^N\exp\left\{\sum_{n=1}^{N}\ln\left[1+\psi(E,B_y)-\cos\left(\frac{2\pi n}{N}\right)\right]\right\}=\\
=(2J)^N\exp\left\{\frac{N}{2\pi}\int\limits_{0}^{2\pi}\ln\left[1+\psi(E,B_y)-\cos\phi\right]{\rm d}\phi\right\}=\\
=J^N\left(\Psi(E,B_y)\right)^N,
\end{multline}
where
\begin{equation}\label{eq:function_Psi}
\Psi(E,B_y)=1+\psi(E,B_y)+\sqrt{\psi(E,B_y)(\psi(E,B_y)+2)}.
\end{equation}
After performing analogous calculations, one can obtain the following result:
\begin{equation}\label{eq:newfactor}
\sqrt{\frac{\det\mathcal{H}[s_{GS}]}{\det\mathcal{H}[s_{LM}]}}=
\left[\frac{\Psi(E,B_y)\Psi(-E,B_y)}{\Psi(E,-B_y)\Psi(-E,-B_y)}\right]^{N/2}.
\end{equation}
For the systems under consideration, $\psi(\pm E,\pm B_y)\ll 1$.
This allows us to simplify Eq.~(\ref{eq:newfactor}) to:
\begin{equation}\label{eq:newfactor1}
\sqrt{\frac{\det\mathcal{H}[s_{GS}]}{\det\mathcal{H}[s_{LM}]}}=
\exp\left[\frac{\mu B_y N}{\sqrt{8JK}}\left(\sqrt{\frac{K}{K+E}}+\sqrt{\frac{K}{K-E}}\right)\right].
\end{equation}
Eq.~(\ref{eq:newfactor1}) exhibits symmetry under the transformation $E\to-E$, which confirms its applicability to both long atomic chains of Co and Fe.

Finally, the frequency prefactors for the transition from the GS to the LM state can be estimated as
\begin{equation}\label{eq:eq:prefactor01}
\nu_{01}=\nu_{01}^{(1)}N_{\rm eff}\exp\left[\frac{\mu B_y N}{\sqrt{8JK}}\left(\sqrt{\frac{K}{K+E}}+\sqrt{\frac{K}{K-E}}\right)\right],
\end{equation}
where $N_{\rm eff}$ is calculated according to Eqs.~(\ref{eq:prefactor02CoI}), (\ref{eq:prefactor02CoII}), or (\ref{eq:prefactor02Fe}).
The resulting dependencies, $\nu_{01}(N)$, are shown in Fig.~\ref{fig5}(c,d) with solid lines.
One can see that the theoretical estimates reproduce the key characteristic of the frequency prefactors $\nu_{01}$ -- an exponential growth with increasing chain length $N$.
In the case of Co chains, the theoretical estimates align well with the numerical results.
However, for Fe chains, the agreement is less satisfactory.
The numerical results are approximately twice as high as the theoretical estimates ($92.8$~THz compared to $49.2$~THz for the Fe chain consisting of 100 atoms).
Nonetheless, this discrepancy is acceptable since we only require agreement in order of magnitude.

\subsection{Magnetization properties}\label{sec:magnetization_properties}

Knowing the energy barriers, we can calculate the magnetization reversal frequencies for the transition from the ground state to the excited state~\cite{9969869}.
For the Co chains, we find that
\begin{equation}\label{eq:nu_up_down}
\nu_{\uparrow\to\downarrow}=\nu_{01}^{I}\exp\left(-\frac{\Delta E_1^I}{k_B T}\right)+\nu_{01}^{II}\exp\left(-\frac{\Delta E_1^{II}}{k_B T}\right),
\end{equation}
for the forward transition, while for the backward transition, we have
\begin{equation}\label{eq:nu_down_up}
\nu_{\downarrow\to\uparrow}=\nu_{02}^{I}\exp\left(-\frac{\Delta E_2^I}{k_B T}\right)+\nu_{02}^{II}\exp\left(-\frac{\Delta E_2^{II}}{k_B T}\right).
\end{equation}
In the case of the Fe chains we should modify the Eqs.~(\ref{eq:nu_up_down}) and (\ref{eq:nu_down_up}) as follows: $\Delta E_1^I=\Delta E_1^{II}=\Delta E_1$, $\Delta E_2^I=\Delta E_2^{II}=\Delta E_2$, $\nu_{01}^I=\nu_{01}^{II}=\nu_{01}$, and $\nu_{02}^I=\nu_{02}^{II}=\nu_{02}$.

\begin{figure}[htb]
\begin{center}
\includegraphics[width=0.95\linewidth]{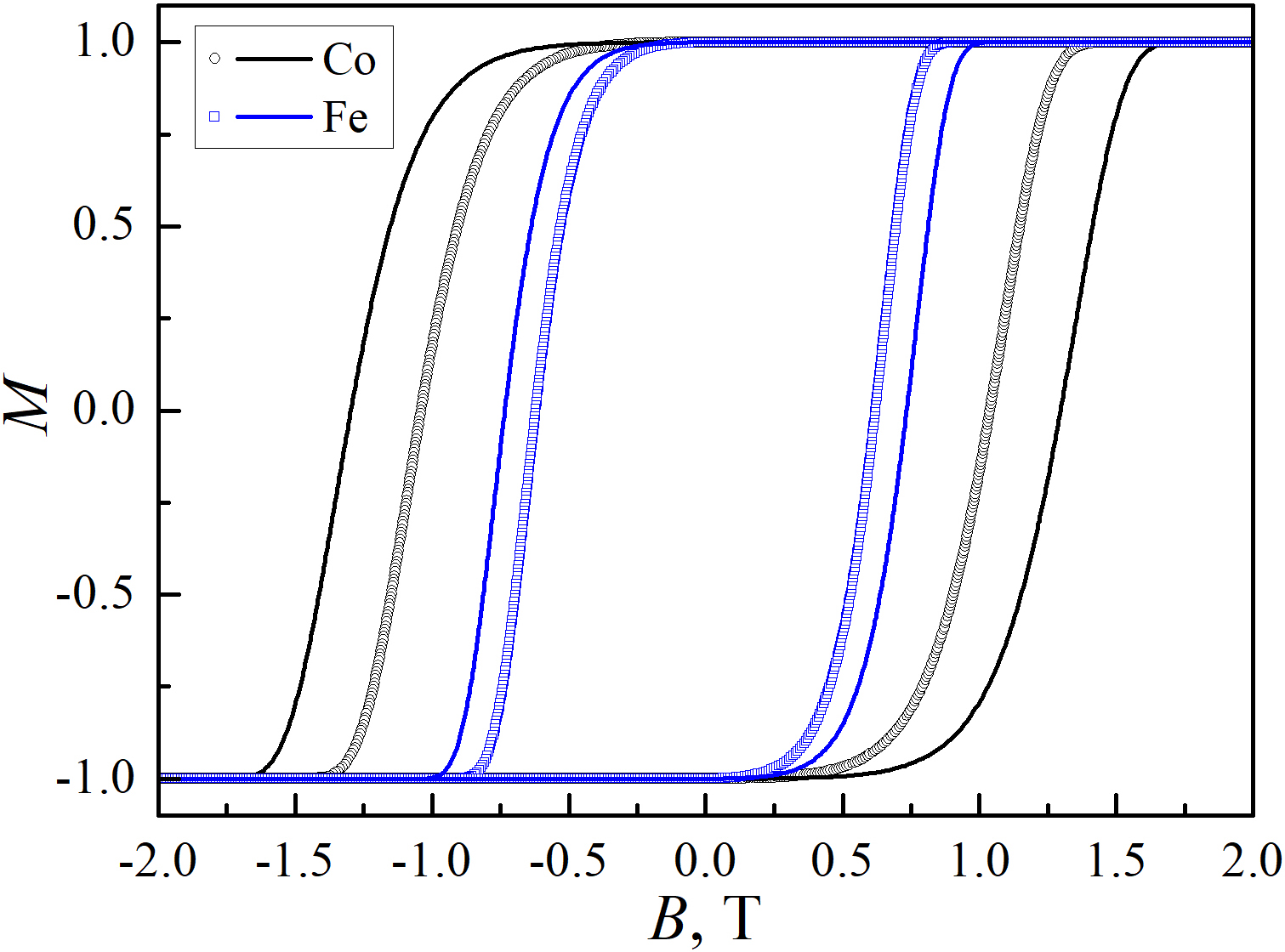}
\caption{\label{fig6} The magnetization curves for the Co and Fe chains on the Pt(332) surface at a temperature of 5~K, with $N=100$, and ${\rm d}B/{\rm d}t=0.04$~T/s.
The points and the solid lines represent the results from the numerical solution of Eq.~(\ref{eq:Meq}), where the energy barriers and frequency prefactors are ether calculated numerically (points), or estimated theoretically (solid lines).}
\end{center}
\end{figure}

It is convenient to measure the magnetization of an atomic chain in dimensionless units, $M\in[-1,1]$.
The magnetization can be determined numerically from the equation~\cite{JETPL103.588,PRB100.224424}
\begin{equation}\label{eq:Meq}
\frac{dM(t)}{dt}=\mathfrak{A}(t)M(t)+\mathfrak{B}(t),
\end{equation}
with the initial condition $M(0)=1$, where $\mathfrak{A}=-\nu_{\uparrow\to\downarrow}-\nu_{\downarrow\to\uparrow}$ and $\mathfrak{B}=\nu_{\downarrow\to\uparrow}-\nu_{\uparrow\to\downarrow}$.
We consider the linear regime of magnetization reversal.
The external magnetic field $B_y$ initially decreases from $B_0=5$~T to $-B_0$ at a constant sweeping rate ${\rm d}B/{\rm d}t$, and then increases from $-B_0$ back to $B_0$ at the same rate.
For numerical calculations, we choose ${\rm d}B/{\rm d}t=0.04$~T/s, which corresponds to a real experiment~\cite{PRB56.2340}.
Figure~\ref{fig6} presents the magnetization curves of the Co and Fe chains, each consisting of 100 atoms on the Pt(332) surface at a temperature of 5~K.
The points in the figure correspond to the energy barriers and frequency prefactors calculated numerically, while the solid lines represent our analytical estimates of these quantities.
It is evident that the magnetization curves based on the analytical estimates are wider than expected.
For Co chains, this discrepancy is attributed to an overestimation of the energy barriers.
For Fe chains, it results from an underestimation of the frequency prefactors.
Let us compare the coercive forces $B_C$ of the atomic chains.
For Co chains, the numerical calculations yield a value of 1.04~T, whereas the analytical estimates lead to the value of 1.30~T.
For Fe chains, the corresponding values are 0.62~T and 0.73~T, respectively.

If a rough estimate of the coercive force is sufficient, it can be obtained without the numerical integration of Eq.~(\ref{eq:Meq}).
Indeed, the coercive force can be qualitatively estimated using the simple equation
\begin{equation}\label{eq:estimation}
\nu_{\downarrow\to\uparrow}(B_c)\approx\frac{{\rm d}B}{{\rm d}t}.
\end{equation}
Applying Eq.~(\ref{eq:estimation}) yields a value of 1.06~T for Co chains and 0.53~T for Fe chains.
It is important to note that this rough estimate tends to underestimate the coercive force.
Note that the mutual compensation of errors can result in the rough estimate aligning more closely with numerical calculations than the solutions derived from Eq.~(\ref{eq:Meq}).
For instance, we observe a notable agreement in the values of $B_C$ for the Co chain, which can be considered a fortunate coincidence.

\begin{figure*}[htb]
\begin{center}
\includegraphics[width=1.0\linewidth]{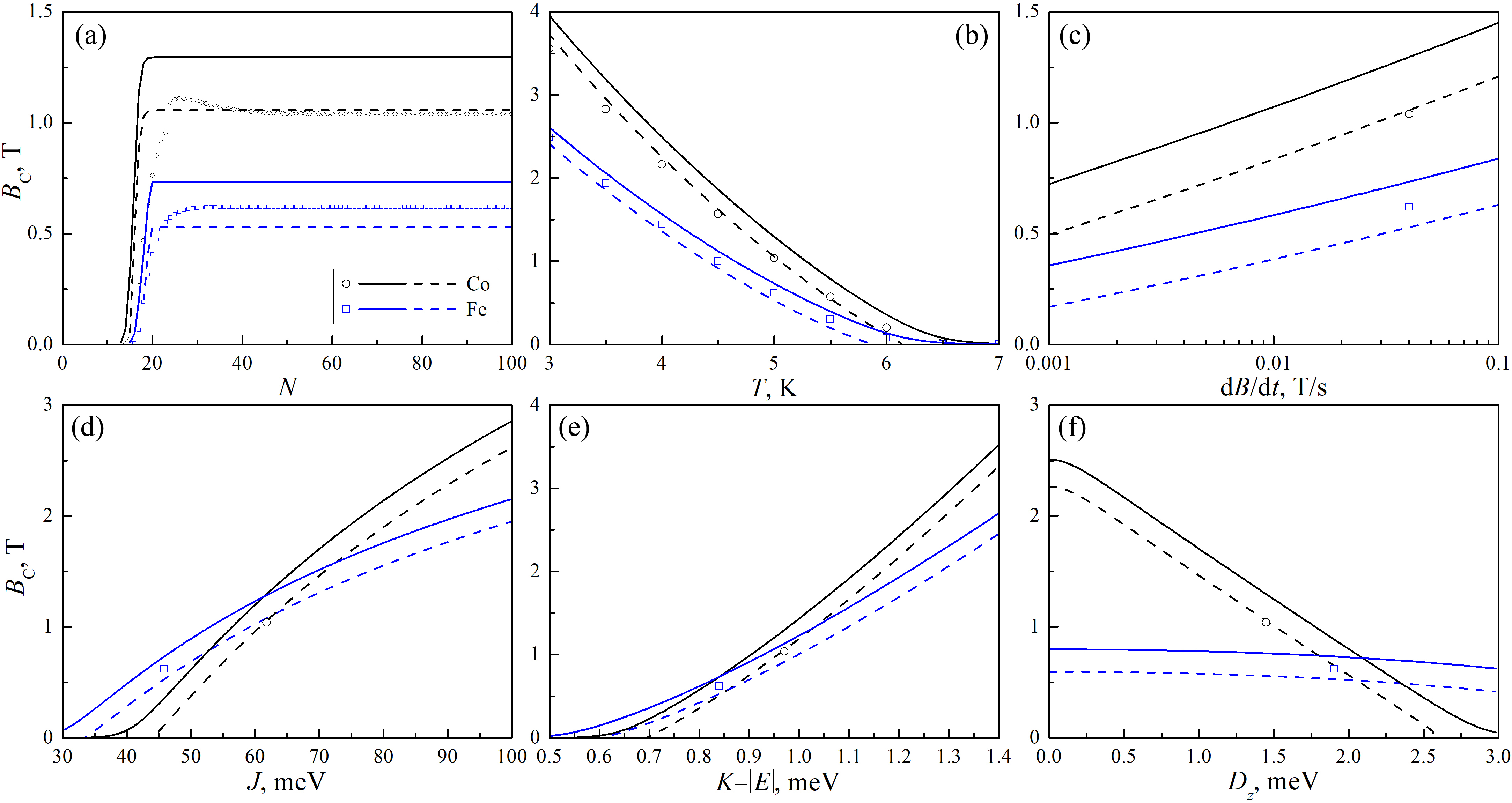}
\caption{\label{fig7}
The dependencies of the coercive force $B_C$ on the following factors are shown: (a) the length of the atomic chain $N$, (b) the temperature $T$, (c) the sweeping rate ${\rm d}B/{\rm d}t$, (d) the exchange integral $J$, (e) the difference in the MAE parameters $K-|E|$, and (f) the DMI parameter $D_z$.
The quantities that are not varying in these dependencies are held constant at the values listed in Table~\ref{table} and $N=100$, $T=5$~K, ${\rm d}B/{\rm d}t=0.04$~T/s.
The points and the solid lines represent the results from the numerical solution of Eq.~(\ref{eq:Meq}), where the energy barriers and frequency prefactors are ether calculated numerically (points), or estimated theoretically (solid lines).
The dashed lines represent rough estimates of $B_C$ according to Eq.~(\ref{eq:estimation}), where both the energy barriers and frequency prefactors are theoretically estimated.
}
\end{center}
\end{figure*}

The analogous calculations of the coercive force can also be performed for shorter Co and Fe atomic chains.
Figure~\ref{fig7}(a) shows the dependencies $B_c(N)$ at $T=5$~K and ${\rm d}B/{\rm d}t=0.04$~T/s.
The points, solid lines, and dashed lines correspond to the three different calculation methods discussed above.
In the case of Fe chains, the functions $\nu_{01}(N)$ and $\nu_{02}(N)$ exhibit discontinuities at $N=16$.
However, the coercive force is negligible at this chain length, resulting in the dependency $B_c(N)$ appearing as a continuous function.
For Co chains, the dependency $B_c(N)$ displays a first-kind discontinuity at $N=24$, which corresponds to discontinuities in the functions $E_1^{II}(N)$ and $E_2^{II}(N)$.
It is noteworthy that the non-monotonic behavior of the frequency prefactors has a minimal effect on the final dependence $B_c(N)$.

The temperature dependencies of the coercive force $B_C$ for Co and Fe chains consisting of 100 atoms are presented in Fig.~\ref{fig7}(b).
The coercive force decreases to zero as the temperature rises from 3~K to 7~K.
Within this temperature range, the coercive force of the Co chain is greater than that of the Fe chain.
The rough estimation of $B_C$ (represented by the dashed lines) aligns more closely with the numerical results (depicted by the points) than the solutions derived from Eq.~(\ref{eq:Meq}) (shown as solid lines).
As previously mentioned, this alignment can be considered a fortunate coincidence.
It is important to note that Eq.~(\ref{eq:estimation}) does not yield a solution at all temperatures.
The maximum temperature corresponds to a zero coercive force, which has the physical meaning of the blocking temperature $T_B$ of the atomic chain.
The blocking temperature is found to be 5.90~K for the Fe chain and 6.14~K for the Co chain.
One can see that the blocking temperature is significantly lower than the critical temperature $T_{\rm cr}$ of the atomic chains, which are approximately $35\pm2$~K for Fe and $40\pm2$~K for Co.
It means that the presented results are consistent with one another.

Figure~\ref{fig7}(c) illustrates the dependencies of the coercive force $B_C$ on the sweeping rate ${\rm d}B/{\rm d}t$.
It is evident that the coercive force is proportional to $\ln({\rm d}B/{\rm d}t)$ over a wide range of ${\rm d}B/{\rm d}t$ values.
Notably, an increase in the sweeping rate by a factor of $k$ corresponds to a decrease of all frequency prefactors by the same factor $k$.
Therefore, we can interpret the presented results as follows: if all frequency prefactors decrease by 100 times, the coercive force approximately doubles.
It is well-known that TST tends to overestimate the frequency prefactors~\cite{RMP62.251}.
According to our TST results (see Fig.~\ref{fig5}(c,d)), the frequency prefactors are on the order of $10^{13}$~Hz.
At the same time, the experimental estimate of the frequency prefactor is $10^9$~Hz~\cite{Nature416.301}, which is $10^4$ times lower than the TST result.
Consequently, our estimates of the coercive force differ by no more than a factor of 4 from those obtained using experimental values of the frequency prefactors (while maintaining the same energy barrier values).
It is worth noting that a linear dependence between the coercive force and the sweeping rate has been reported previously~\cite{PRB73.174418}.
We emphasize that there is no contradiction between this finding and our calculations.
Indeed, the range of ${\rm d}B/{\rm d}t$ values considered in Ref.~\cite{PRB73.174418} is quite narrow, and the logarithmic dependence was mistakenly interpreted as linear.

Now, let us examine how the coercive force depends on the parameters of the Hamiltonian~(\ref{eq:eff_ham}).
We consider chains consisting of 100 atoms at a temperature of 5~K.
The dependencies of the coercive force $B_C$ on the exchange integral $J$ are shown in Fig.~\ref{fig7}(d), with all other parameters taken from Table~\ref{table}.
The values of $B_C$ corresponding to the $J$ values listed in Table~\ref{table} are represented by points on the graph.
We see that the coercive force increases monotonically with an increase in $J$.
Interestingly, if the exchange integral $J$ were identical for both Co and Fe chains, the coercive force of the Fe chains would exceed that of the Co chains for $J<60$~meV, while the opposite would be true for $J>60$~meV.

Figure~\ref{fig7}(e) presents the dependencies of the coercive force on the difference between the MAE parameters $K-|E|$.
All other parameters are taken from Table~\ref{table}.
The values of $B_C$ corresponding to the $K-|E|$ values listed in Table~\ref{table} are represented by points on the graph.
Strictly speaking, the coercive force depends on the parameters $K$ and $|E|$ separately.
However, the dependencies $B_c(K-|E|)$ at fixed $K$ and fixed $|E|$ are nearly identical, with no visual differences evident in Fig.~\ref{fig7}(e).
It is clear that the coercive force increases monotonically with an increase in $K-|E|$.
Note that if the value of $K-|E|$ were the same for both Co and Fe chains, the coercive force of the Fe chains would be greater than that of the Co chains for $K-|E|<0.85$~meV, while the opposite would hold true for $K-|E|>0.85$~meV.

Finally, let us examine the dependencies of the coercive force on the value of $D_z$, with all other parameters taken from Table~\ref{table}.
The values of $B_C$ corresponding to the $D_z$ values listed in Table~\ref{table} are represented by points on the graph.
One can see that the coercive force decreases monotonically with an increase in $D_z$.
In other words, the DMI contributes to a reduction in the coercive force of the chains.
This effect is particularly pronounced in the case of Co chains, while it is relatively weak for Fe chains, attributed to the different structures of the domain walls.
Notably, if the value of $D_z$ were the same for both Co and Fe chains, the coercive force of the Co chains would exceed that of the Fe chains for $D_z<2.1$~meV, whereas the opposite would be true for $D_z>2.1$~meV.

\section{Conclusion}\label{sec:conclusion}

In summary, we have investigated the different mechanisms underlying the magnetization reversal of finite-length Co and Fe chains on the Pt(332) surface.
For short atomic chains, the magnetization reversal occurs through the simultaneous reversal of all magnetic moments.
In contrast, the magnetization reversal of long atomic chains is facilitated by the formation of domain walls, which exhibit distinct structures for Co and Fe chains.
Utilizing the GNEB method, we have determined the energy barriers associated with the magnetization reversal of chains consisting of 5 to 100 atoms.
Additionally, we calculated the frequency prefactors within the framework of the harmonic approximation of TST.
The dependencies of these prefactors on both the length of the chain and the external magnetic field are significant and exhibit non-monotonic behavior.

We have proposed a theoretical approach that qualitatively describes the numerical dependencies for the energy barriers and frequency prefactors associated with magnetization reversal.
The magnetization curves derived from our theoretical estimates show qualitative agreement with those obtained from numerical calculations.
The developed analytical framework enables us to estimate the coercive force of atomic chains across a broad range of lengths, temperatures, sweeping rates, and parameters of the effective Hamiltonian~(\ref{eq:eff_ham}).
This indicates that our theoretical approach is applicable not only to the Co and Fe chains on the Pt(332) surface but also to a wide class of one-dimensional magnetic systems that can be characterized by the Hamiltonian~(\ref{eq:eff_ham}).

It is important to note that the coercive force of atomic chains is significantly influenced by the parameters $J$, $K$, $E$, and $D_z$ of the model. 
By comparing the theoretical estimates of the coercive force with experimental data (such as those presented in~\cite{Nature416.301}), we can identify limitations within the parametric space of the model\footnote{Unfortunately, there is currently a scarcity of experimental data on Co/Pt and Fe/Pt systems.}.
Consequently, the analytical approach we have presented paves the way for determining the parameters of the effective Hamiltonian~(\ref{eq:eff_ham}) directly from experimental observations.

Finally, we must emphasize that calculations based on TST tend to yield overestimated values for the frequency prefactors~\cite{RMP62.251}.
The next step in advancing the theory of magnetic properties in finite-length atomic chains should involve calculating the frequency prefactors using a more precise reaction rate theory \cite{RMP62.251, Coffey.ch5}.
We intend to focus our future works on addressing this issue.

\section*{Acknowledgements}
The research is carried out using the equipment of the shared research facilities of HPC computing resources at Lomonosov Moscow State University~\cite{NIVC1,NIVC2}.
E.~S.~Glazova acknowledges the financial support of the Theoretical Physics and Mathematics Advancement Foundation ``BASIS''.

\bibliography{Co_Fe_field.bib}

\end{document}